\documentclass[preprint,pre,aps]{revtex4}
\usepackage[pdftex]{graphicx} 
\begin{document}
 \title{Patterns with long and short-range order in monoloyers of binary mixtures with competing interactions }
 \author{ M. Litniewski}
\affiliation{Institute of Physical Chemistry,  Polish Academy of Sciences, 01-224 Warszawa, Poland, }  
\author{ W. T. Gozdz}
\affiliation{Institute of Physical Chemistry,  Polish Academy of Sciences, 01-224 Warszawa, Poland, } 
\author{  A. Ciach\footnote{ the email address of the corresponding author:  alina.ciach@gmail.com }}
\affiliation{Institute of Physical Chemistry,  Polish Academy of Sciences, 01-224 Warszawa, Poland, }
 \date{\today}

 \begin{abstract}
 Lateral microsegregation in a monolayer of a binary mixture of particles or macromolecules is studied by MD simulations in a generic model with the interacting potentials inspired by effective interactions in biological or soft-matter systems. In the model, the energy is minimized when like particles form small clusters, and the cross-interaction is of opposite sign. We show that the laterally microsegregated components in the dense ordered phases form alternating stripes for similar densities, or the clusters of the minority component fill the hexagonally distributed voids formed in the dense phase of the majority component. A qualitative phase diagram in the plane of densities of the two components is constructed for low temperatures. An addition of the second component significantly enlarges the temperature range of the stability of the ordered phases compared to the stability of these phases in the one-component system. At higher temperatures, the disordered phase consisting of individual particles, one-component clusters and two-component super-clusters of various sizes is stable. The product kn(k), with n(k) denoting the average number of super-clusters composed of k particles, decays exponentially with k, and the inverse decay rate depends linearly on temperature. 
 \end{abstract} 
 \maketitle
 
 \section{introduction}
 Complex solvents in soft- and living matter often induce complex interactions between particles or macromolecules.
 A notable example
 is the thermodynamic Casimir potential between objects 
present in binary or multicomponent solvents that are close to a miscibility critical point~\cite{hertlein:08:0}. The objects attracting the same or different  component of a solvent attract or repell each other,  
respectively, and the range of these interactions is equal to the correlation length of the concentration fluctuations in the solvent. 
When the macromolecules or particles are charged, electrostatic interactions with a sign opposite to the sign of the Casimir potential appear~\cite{hertlein:08:0,gambassi:09:0,pousaneh:12:0}.  When the attraction and the repulsion of any origin are both present and dominate at different separations between the particles or macromolecules, complex patterns may appear~\cite{stradner:04:0,campbell:05:0,archer:08:0,ciach:13:0,
almarza:14:0,marlot:20:0,litniewski:21:0, zhuang:16:0,lindquist:16:0,pini:17:0,rey:20:0,grishina:20:0,rauh:17:0}.

 Interestingly, multicomponent biological membranes in homothermic living organisms are close to the miscibility critical point~\cite{veatch:07:0,machta:12:0,shaw:21:0}. It was hypothesised that the Casimir attraction may induce aggregation of the membrane proteins, but a macroscopic separation of the proteins was not observed~\cite{shaw:21:0,rozovsky:04:0}. The membrane proteins are charged and according to recent discoveries, the screening length can be large in ionic solutions as concentrated as in living organisms~\cite{smith:16:0,lee:17:0,groves:24:0,Zeman2021,kumar:22:0,
 safran:23:0,ciach:23:0,Yang2023}. When the screening length is larger than the correlation length but the strength of the screened electrostatic interactions is smaller than the strength of the Casimir potential, then the interaction between charged particles or macromolecules is attractive at short and repulsive at large distances (SALR).
  SALR interactions would lead to formation of  aggregates of the membrane proteins of a  size determined by the range of the attraction, and separated by a distance determined by the range of the repulsion. Small, well separated clusters of membrane proteins are indeed observed.
  
The  hypothesis that like membrane inclusions interact with the  SALR potential, and that the interaction between oppositely charged inclusions favouring different components of the lipid bilayer is repulsive at short and attractive at large distances was not verified yet. It inspires, however, a general question of pattern formation in monolayers of binary mixtures of particles with such interactions.
We try to answer this question by considering models with different ranges and shapes of the attractive and repulsive parts of the potentials~\cite{patsahan:21:0,patsahan:24:0,ciach:23:2,virgiliis:24:1,virgiliis:24:0}.  

Although the models with the above interactions are inspired by the interactions between two types of membrane inclusions, one could design soft-matter systems that mimic  such type of interactions between particles.  Colloidal particles interacting with the thermodynamic Casimir potential have been already studied experimentally and theoretically~\cite{nguyen:13:0,vasiliev:21:0,marino:21:0,
marlot:19:0,marlot:20:0}. Moreover, different attractive forces, such as depletion or capillary interactions can be present
~\cite{roth:00:0,MAO199510,campbell:05:0,KRALCHEVSKY2000145,
rey:20:0,grishina:20:0}, 
and repulsion may follow from dipole-dipole interactions or soft polymeric shells~\cite{marlot:19:0,rauh:17:0,rey:20:0}. Thus, one can design many soft-matter systems with competing attractive and repulsive potentials.
Spontaneous patterns at different length scales may find various applications, so this problem is of practical significance as well. 

 Mixtures of self-assembling particles
attract increasing attention~\cite{sweatman:18:0,sweatman:21:0,munao:22:0,
prestipino:23:0,munao:23:0,ciach:23:2,virgiliis:24:0,virgiliis:24:1}, but 
a complete phase diagram has not been determined yet for any particular system. In the case of the SALR interactions between like particles and cross-interaction of the opposite sign,
preliminary results have been obtained  for intermediate temperatures for interactions leading to large aggregates (tens of particles) in mean-field (MF) approximation and by MC simulations~\cite{patsahan:24:0}. The energetically favoured patterns on the plane of chemical potentials were determined for a potential favoring small aggregates on a triangular lattice by direct calculations and MC simulations~\cite{virgiliis:24:1}. The question how the structure of the ordered phases  and transitions between them depend on the shape of the interactions (determining the size of the aggregates), temperature, density  and composition, however, remains to a large extent open. 

By comparing the snapshots in the one-component SALR model and in the considered mixture, one can see that the structure in the dilute disordered phase in the two cases is significantly different. Dispersed small one-component clusters in the first system are replaced by much larger super-clusters made of alternating clusters of the two components in the second one~\cite{Ciach:20:1,ciach:23:2,virgiliis:24:0}.  However,
quantitative characterization of the   patterns lacking long-range order in the self-assembling mixture  remains a real challenge.

In this work we focus on the low-temperature ordered phases with laterally microsegregated components in the monolayer of particles with energetically favoured alternating thin stripes of the two components. We choose interactions that favour the stripes consisting of two adjacent chains of  particles of the same component (a bilayer in 2D). We perform MD simulations, and construct a qualitative
phase diagram with a topology that should be common for many mixtures with interactions favouring alternating thin stripes of the two components.  
 In addition, we focus on the high-temperature region of the phase diagram where the disordered phase is stable, and study structure of this phase by calculating the size distribution of self-assembled clusters and super-clusters. We
 investigate the evolution of the structure upon increasing temperature, density and mole fraction of the particles.
 
 In sec.~\ref{sec:model}, we introduce the model.
  In sec.~\ref{sec:theory}, we summarize the MF results for the $\lambda$-surface separating the low- and high temperature regions in the phase space, where 
  the phases with the long-range order are and are not expected, respectively. The simulation method is described in sec.~\ref{sec:simulations}. Our results for the patterns formed spontaneously at low $T$ are presented in sec.~\ref{sec:results1}, and for the structure of the disordered phase in sec.~\ref{sec:results2}.  The last section contains summary and  concluding remarks.
 
 \section{The model}
 \label{sec:model}
 We assume that the particles of both components have  hard or nearly hard spherical cores, and the  sizes of the cores are equal.   As in the previous studies~\cite{patsahan:21:0,litniewski:21:0}, we assume the same interactions between like particles, 
 \begin{equation}
 \label{uii}
 u_{11}(r)=u_{22}(r)=u_{hc}(r)+u(r),
\end{equation}
whereas for the cross-interaction we assume 
\begin{equation}
\label{u12}
u_{12}(r)=u_{hc}(r)-u(r).
\end{equation}
We are interested in self-assembly into bilayers or small clusters in the absence of the second component, therefore we assume the shape of the like-particles interactions with a deep minimum for a relatively small distance. The potential satisfying the above requirements has in particular the form
 \begin{equation}
 \label{u}
u(r)=\frac{-2.725}{r^6} + 1.5\frac{\exp(-r/2)}{r}.
 \end{equation}
 
In the mesoscopic theory, $u_{hc}$ is the hard-sphere potential, $u_{hc}^{t}(r)=\infty$ for $r<1$  and $u_{hc}^{t}(r)=0$ for $r>1$ that prevents from overlapping of the particle cores.
  To mimic the (nearly) hard cores in the MD simulations, we assume very strong repulsion for $r<1$ of the form
 \begin{equation}
 \label{uhc}
 u_{hc}^s(r)=\frac{2.725}{r^{30}}.
 \end{equation}
In the above equations, the length is in units of the particle diameter $\sigma$, and the energy is in arbitrary units $\epsilon$. 

\begin{figure}
 \includegraphics[scale=0.4]{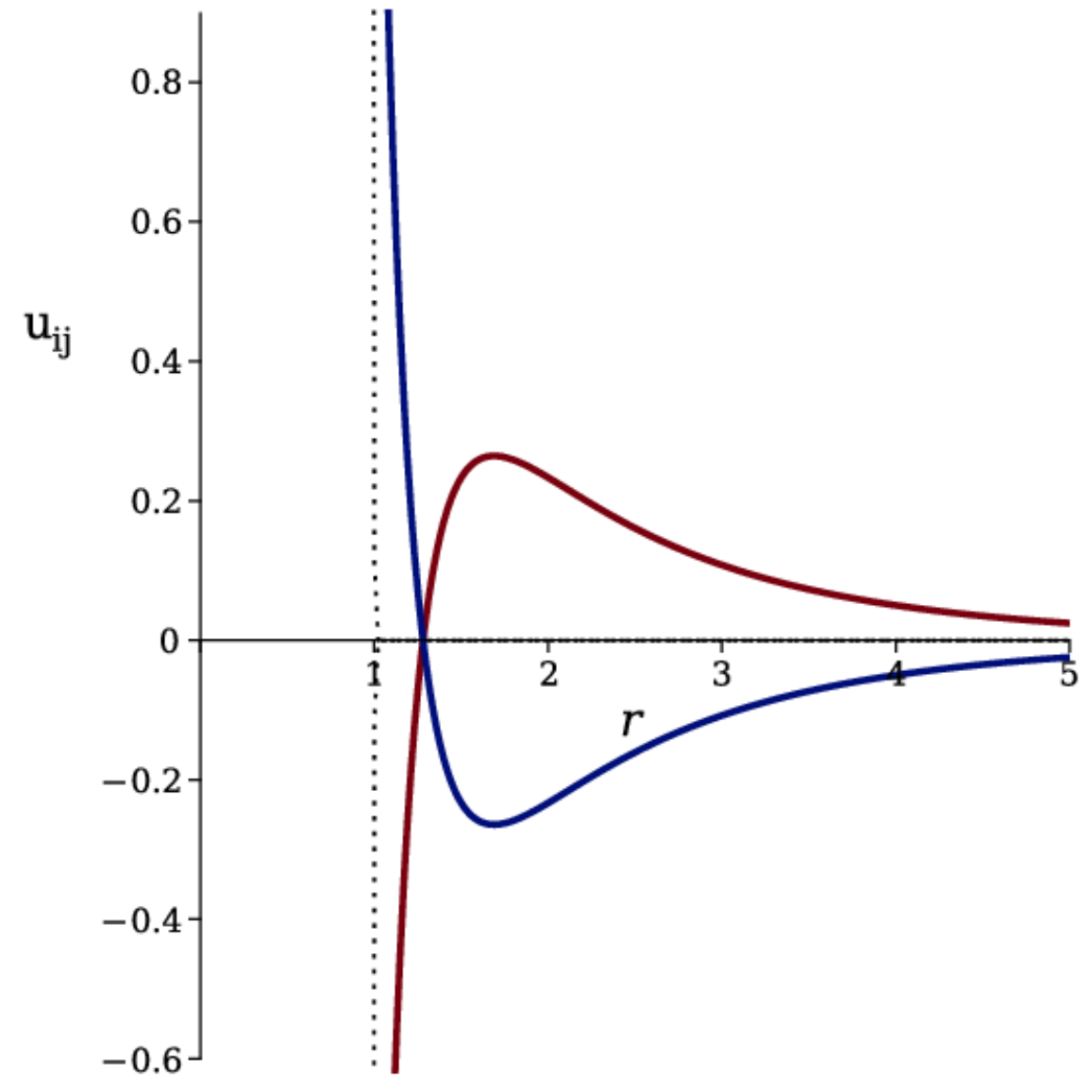} 
 \includegraphics[scale=0.4]{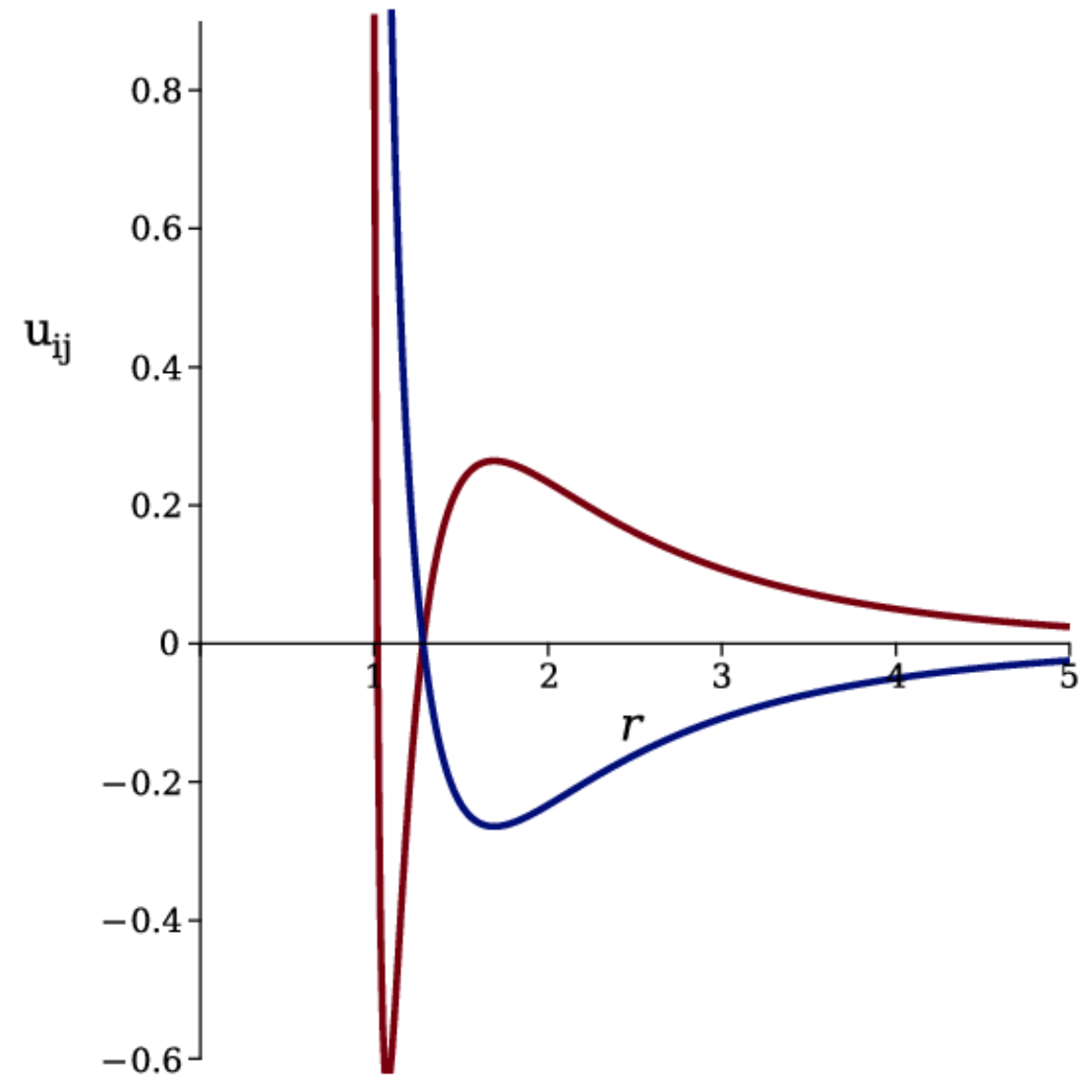}
 \caption{The interaction potential $u_{ij}$  with $i,j=1,2$ between like particles, $u_{ii}(r)$ (Eqs.(\ref{uii}) and (\ref{u}), red  line)  and the cross-interaction $u_{12}(r)$ (Eq.(\ref{u12}) and (\ref{u}),  blue line). Left panel: hard core with diameter $\sigma=1$ (dotted line), $u_{hs}=u_{hs}^t$. Right panel: nearly hard core, $u_{hs}=u_{hs}^s$ (Eq. (\ref{uhc})), as in the simulations.}
 \label{f1}
 \end{figure}

To compare the theoretical results obtained for $u_{hc}^{t}(r)$ with the MD simulations obtained with $u_{hc}^s(r)$ given by Eq.(\ref{uhc}), we should take into account that the minimum of  $u_{ii}(r)$ depends on the form of $u_{hc}$.  $u_{ii}(r)$ takes the minimum at $r=1$, and $u_{min}^t=u(1)=-1.815$ when $u_{hc}=u_{hc}^t$, and $u_{ii}(r) $ crosses zero for $r\approx 1.2808$. For $u_{hc}$ approximated by Eq.(\ref{uhc}),  the minimum of $u_{ii}(r)$ is $u_{min}=u_{ii}(1.075)\approx-0.6393$, and $u_{ii}(r) $ crosses zero for $r\approx 1.28$. 
We assume that the appropriate energy scale is the minimum of $u_{ii}$, and introduce the dimensionless temperature $T^*=k_BT/|u_{min}|$, with $k_B$ the Boltzmann constant and $u_{min}=u_{min}^s\approx -0.6393$ in the simulations and   $u_{min}=u_{min}^t\approx -1.815$ in the theory. 
    The  potentials $u_{ii}$ for $i=1,2$ and $u_{12}$ are shown in Fig.~\ref{f1}.
 
 \section{Stability of the disordered phase in mean-field approximation}
\label{sec:theory}
  In theoretical studies, the stability analysis of the disordered phase in MF gives the first approximate information about the onset of the periodic ordering. 
   In the disordered phase, the average density of both components is position independent. This phase can be inhomogeneous when the  self-assembled aggragates are mobile. It looses stability when the second functional derivative of the grand potential functional of  the local 
   densities
   $\rho_1({\bf r}),\rho_2({\bf r})$,
   \begin{equation}
   \label{Omega}
   \Omega[\rho_1,\rho_2]=U[\rho_1,\rho_2]-TS[\rho_1,\rho_2]-\mu_1\int d{\bf r}\rho_1({\bf r})-\mu_2\int d{\bf r}\rho_2({\bf r})
\end{equation}  
  is no longer positive definite for constant functions $\rho_1,\rho_2$. 
The internal energy for the interactions having the property  $u_{11}=u_{22}=-u_{12}$ outside the hard cores has the form
  \begin{equation}
 \label{U}
 U[\rho_1,\rho_2]=\frac{1}{2}\int d{\bf r}_1\int d{\bf r} c({\bf r}_1)V(r)c({\bf r}_1+{\bf r})=\frac{1}{2}\int d{\bf k} \hat c({\bf k})\hat V(k) \hat c(-{\bf k}),
 \end{equation}
where $c=\rho_1-\rho_2$, 
$r=|{\bf r}|, k=|{\bf k}|$ and  $V(r)=u(r)g(r)$, where $g(r)$ is the pair distribution function.   $\hat V(k)$ is  the Fourier transform of  $V(r)$.
In the case of hard cores, $g(r)=0$ for $r<1$. We neglect correlations outside the cores in MF, therefore we make the approximation $g(r)=\theta(r-1)$.  The entropic contribution in (\ref{Omega}) is approximated by the free energy of a hard-sphere mixture in the local  density approximation,
\begin{equation}
-TS=\int d{\bf r}[k_BT(\rho_1\ln\rho_1+\rho_2\ln\rho_2)+ f_{ex}(\rho)],
\end{equation}
 where $\rho=\rho_1+\rho_2$ and the hard-spheres packing is described by the free-energy density $f_{ex}$ in the Carnahan-Starling approximation~\cite{patsahan:24:0}. 

\begin{figure}
 \includegraphics[scale=0.35]{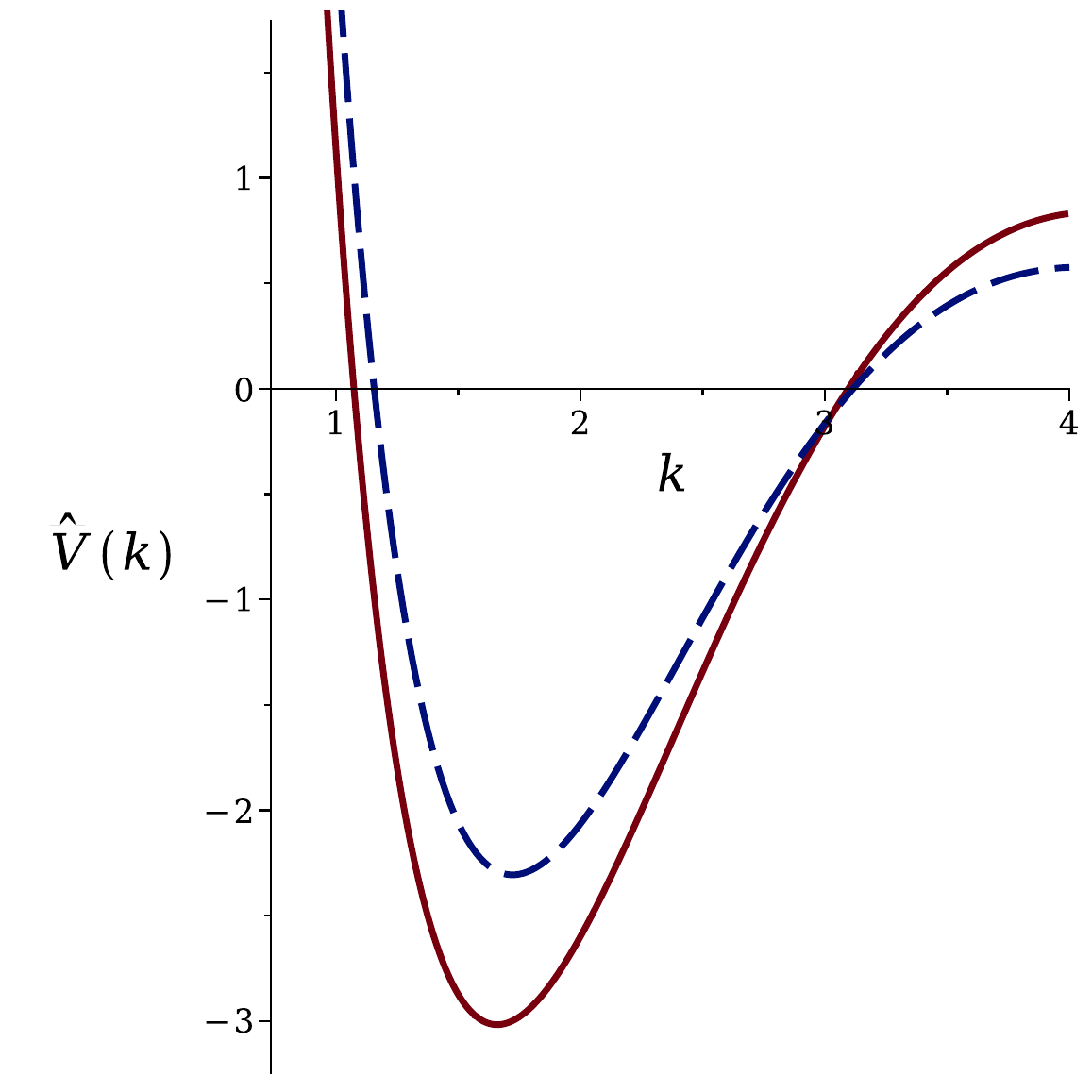} 
 \caption{The  interaction potential times the pair distribution  function in Fourier representation, $\hat V(k)$. Red solid line:  $u_{hs}=u_{hs}^t$  (hard spheres). Blue dashed line: $u_{hs}=u_{hs}^s$ (Eq.(\ref{uhc}), nearly hard spheres). }
 \label{f2}
 \end{figure}

 The period of the  self-assembled patterns should correspond to the densities $\rho_i({\bf r})$ that minimize the internal energy.
  The wave number of the energetically favored density oscillations corresponds to the minimum of $\hat V(k)$ at $k=k_0$. In our model 
  $k_0\approx 1.6558$ giving the period of concentration oscillations $2\pi/k_0\approx 3.8$, consistent with alternating bilayers of the particles of the two components.
   $\hat V(k)$ for our model is shown in Fig.\ref{f2}. 
 clusters of different components are

 \begin{figure}
 \includegraphics[scale=0.7]{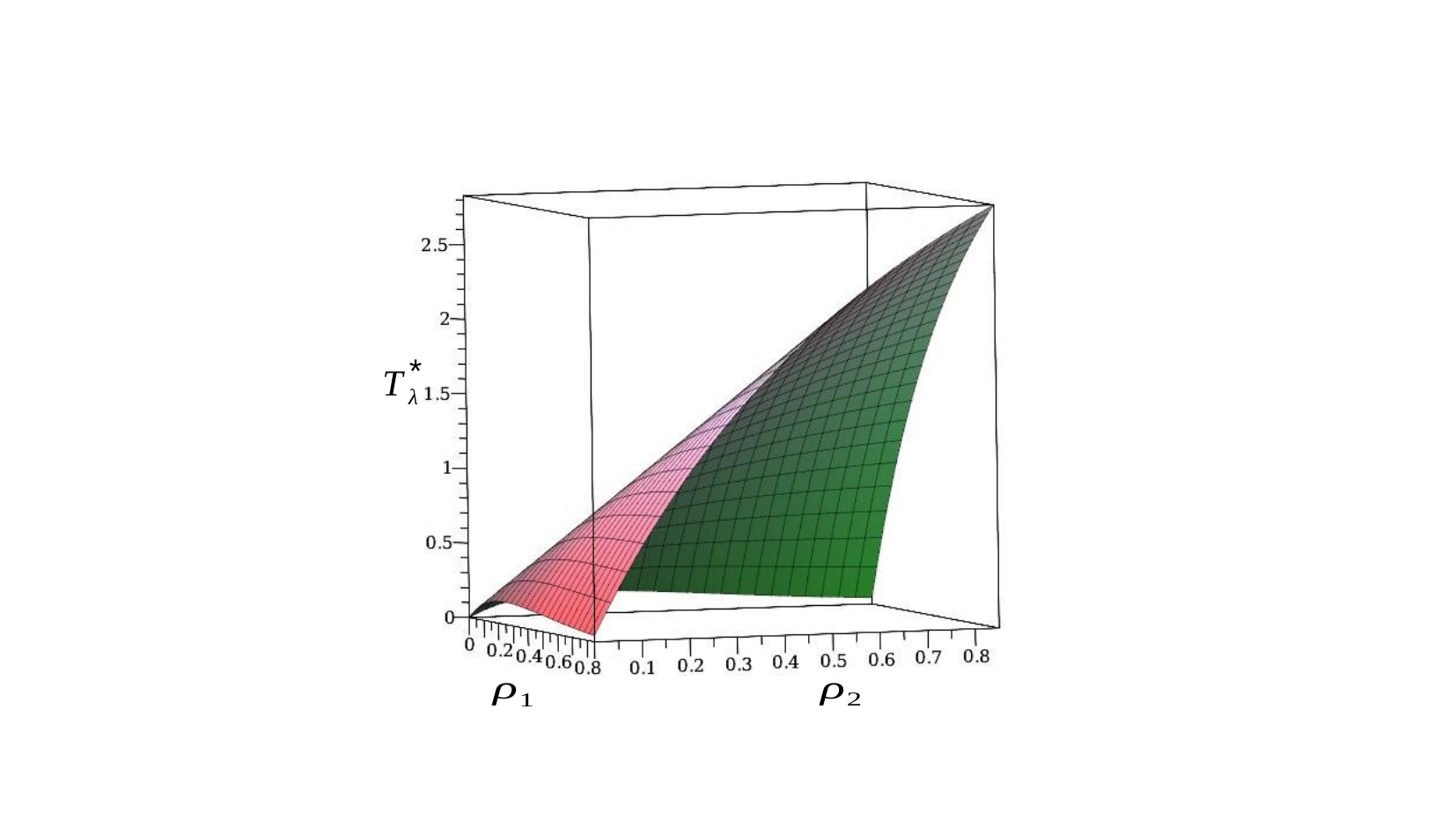}
 \caption{The $\lambda$-surface of the MF boundary of stability of the disordered phase (Eq.~(\ref{Tlambda})). On the low-$T$ side, the disordered phase is unstable in the MF approximation, and self-assembly into alternating aggregates of the particles of the two components takes place in typical configurations. Ordered patterns can be stable on this side of the $\lambda$-surface. Number densities $\rho_i$ are dimensionless, $T^*=kT/|u_{min}|$. }
 \label{f3}
 \end{figure}
   The boundary of stability of the disordered phase in MF, $T_{\lambda}(\rho_1,\rho_2)$, is called $\lambda$-surface, and was determined in Ref.~\cite{patsahan:24:0}. In our temperature units it has the form:
    \begin{equation}
T^*_{\lambda}(\bar\rho_1,\bar\rho_2)=\frac{\big[\bar\rho+(\bar\rho^2-\bar c^2)A_2(\bar\rho)\big]}{1+\bar\rho A_2(\bar\rho)}\frac{\hat V(k_0)}{u_{min}^t},
\label{Tlambda}
 \end{equation}
 where $\bar c=\bar\rho_1-\bar\rho_2$ and $\bar \rho=\bar\rho_1+\bar\rho_2$ are the equilibrium values of the concentration $c$ and the density of the particles $\rho$ in the disordered phase, and
  \begin{equation}
\label{An}
A_2(\rho)=\frac{1}{k_BT}\frac{d^2 f_{ex}(\rho)}{d \rho^2}.
\end{equation} 
   
$T^*_{\lambda}(\bar\rho_1,\bar\rho_2)$ separates the phase space $(\bar\rho_1,\bar\rho_2,T^*)$ into the high-$T$ region where in MF the disordered phase is stable or metastable, and the low- $T$  region where the disordered phase is unstable. In fact local fluctuations $c({\bf r})$ restore stability of the disordered phase beyond MF~\cite{patsahan:21:0}, and the  $\lambda$-surface separates the phase space regions with dominating homogeneous and inhomogeneous distributions of the particles in instantaneous states. The ordered periodic phases can be expected on the low-$T$ side of
 the $\lambda$-surface shown for our model in Fig.~\ref{f3}.
 
 As already noted in Ref.~\cite{patsahan:24:0}, the periodic order can be present at a temperature incresing linearly with $\rho$ for $\rho_1\approx\rho_2$  (Fig.~\ref{f3} and Eq.~(\ref{Tlambda})). For increasing difference in densities of the two components, however, the temperature at the boundary of stability of the disordered phase rapidly decreases to much lower values. This means that the ordered periodic patterns in the one component system or with a small addition of the minority component can be expected only for much lower temperatures.   
 Note that high or low temperature refers to $T^*=k_BT/|u_{min}|$, i.e it strongly depends on the strength of the interactions. 
 
 In Ref.~\cite{patsahan:24:0}, the phase diagram of a similar mixture, but with much larger period of concentration oscillations was determined under the assumption that $c({\bf r})$ is a superposition of sinusoidal waves with the period $2\pi/k_0$  in different directions. Such assumption for ensemble averaged concentration can be valid for thick layers of particles and for not too low temperature~\cite{pini:17:0}. For the stripes made of  two adjacent chains of the same particles, however, large deviations of the concentration profiles from the sinusoidal shape are expected in the ordered phases. For this reason, the results of Ref.~\cite{patsahan:24:0} do not apply to our model, especially at low temperature. 
 
 \section{simulation method}
\label{sec:simulations}
   We are interested in spontaneous pattern formation in monolayers of particles or macromolecules anchored in lipid bilayers or adsorbed at interfaces, where small out of plane displacements are possible. To mimic such situation in simulations, we consider particles in a rectangular box $L_x\times L_y\times L_z$ with $L_x= L_y=L$ and with periodic boundary conditions along these two directions, and $L_z=2.8$. 
Motion of the particles along the z axis is additionally strongly restricted by the external potential 
 \begin{equation}
 \label{uz}
u_z(z)=\frac{1}{z^{30}}-\frac{1}{z^{6}}+\frac{1}{(z-2.8)^{12}}.
\end{equation}

At $z = 1.07$, the  potential $u_z(z)$ has a deep minimum, $u_z(1.07)=-0.84$  (in $\epsilon$ units). As a consequence,  in the considered temperature range
 the equilibrium states attained by the
systems may be treated as two dimensional ones. Our simulations were performed applying
standard molecular dynamics method~\cite{allen:90:0}. The system temperature was kept constant by scaling
particle velocities. The simulation procedure is described in detail in Ref.~\cite{patsahan:21:0,litniewski:21:0}.

 We started the initial simulations  at high temperature with particles distributed randomly in the simulation box, and with $N$ and $L$ chosen such that the two-dimensional density $\rho=N/L^2$ was somewhat smaller than the density expected for the crystal. During the simulations, the temperature was gradually decreased until 
 a  monocrystal with the two-dimensional density $\rho_s$ larger than $N/L^2$ was formed. The obtained monocrystal occupied a part of the simulation box, and the remaining part of the box was esentially empty. 
 
 For studies  of two-phase systems we chose the final state of the  simulations with $N_1=N_2=1800$ and $L=84$ (giving $N/L^2=0.51$), where the solid phase had the characteristic pattern with the repeated motif of two adjacent chains of particles of the first component followed by two such chains of particles of the second component, and its density was $\rho=0.76$.
Further simulations started from this configuration, but with
 the size of the simulation box  increased from $L=84$ to $L=180$.
 Next  the particles of the second type were added gradually to the system in the way described in Ref.~\cite{litniewski:21:0}, with the temperature fixed at $T^* = 0.156$. In this way, a few different systems were obtained with $N_1=1800$,
and with $N=N_1+N_2$  in the range $3840<N<17600$. The simulations of each 
  system created in this way were performed at $T^* = 0.156$ and with fixed $N_1, N_2$. 
In the additional simulations, the temperature was increased until the one-phase systems were obtained. For some systems, the simulations were carried out at the temperature up   to $T^* = 0.469$.

       In order to study the structure of the disordered phase, we considered the one-component clusters in the same way as in the one-component SALR~\cite{litniewski:19:0}, and introduced the concept of two-component super-clusters. A cluster is a compact aggregate  of  particles;   a particle belongs to a cluster
when its distance from at least one particle in the cluster is smaller than $r_{cl} = 1.2835$, where $r_{cl}$ is the distance for which $u_{ii}(r)$  (Eq.~(\ref{uii})) changes sign ($u_{ii}(r_{cl})=0$). The super-cluster is a group of connected clusters of different components.  Because the minimum of $u_{12}(r)$ is very flat and the range of the interactions is large, there is no unique way of defining the connectivity between clusters of different components. We  arbitrarily choose the distance  $r_{scl} = 2.15$ between particles belonging to different clusters as the upper limit for formation of a bond between these clusters. We verified that different reasonable values of  $r_{scl}$ do not influence the results on the  qualitative level.    According to Figure~\ref{f1}, $r_{scl}$ is significantly larger than the distance corresponding to the energy minimum for $u_{12}$. On the other hand,
$|u_{12}(r_{scl})|$ is only slightly lower than the minimum value. Cluster and super-cluster size distribution was obtained by averaging over at least $10^5$ configurations. 

We compared two connectivity criteria for super-cluster formation. In the first one, one bond between particles belonging to different clusters was sufficient, and in the second one, formation of at least four bonds was neccessary to classify the two considered clusters as a part of a super-cluster. In the case of at least four-bonds formation, the bonding energy between two clusters is larger than $1.4$ (in $\epsilon$ units). A disadvantage of the first criterion was formation of transient bonds during time evolution and short life-time of the super-cluster consisting of two weakly connected parts. We choose for the analysis of the structure of the disordered phase the second criterion. We verified that there was no qualitative difference between the results obtained with the two connectivity criteria. A quantitative difference, however, was  significant.

    \section{results}
\subsection{Low temperature phase behavior}
\label{sec:results1}
Our aim is a  construction of a qualitative phase diagram with a topology common for various mixtures self-assembling into alternating stripes or other aggregates, rather than 
 determination of a precise phase diagram for our particular form of the interactions.  
 Here we focus on low $T$, where the patterns with a long-range order can appear.
 To study the low temperature behavior in the simulations, we consider $T^*=0.156$ that is close to the $\lambda$-surface for $\rho_1/\rho_2\ll 1$ (Fig.~\ref{f3}), i.e ordered phases in the one-component limit can be stable.

\begin{figure}
 \includegraphics[scale=0.9]{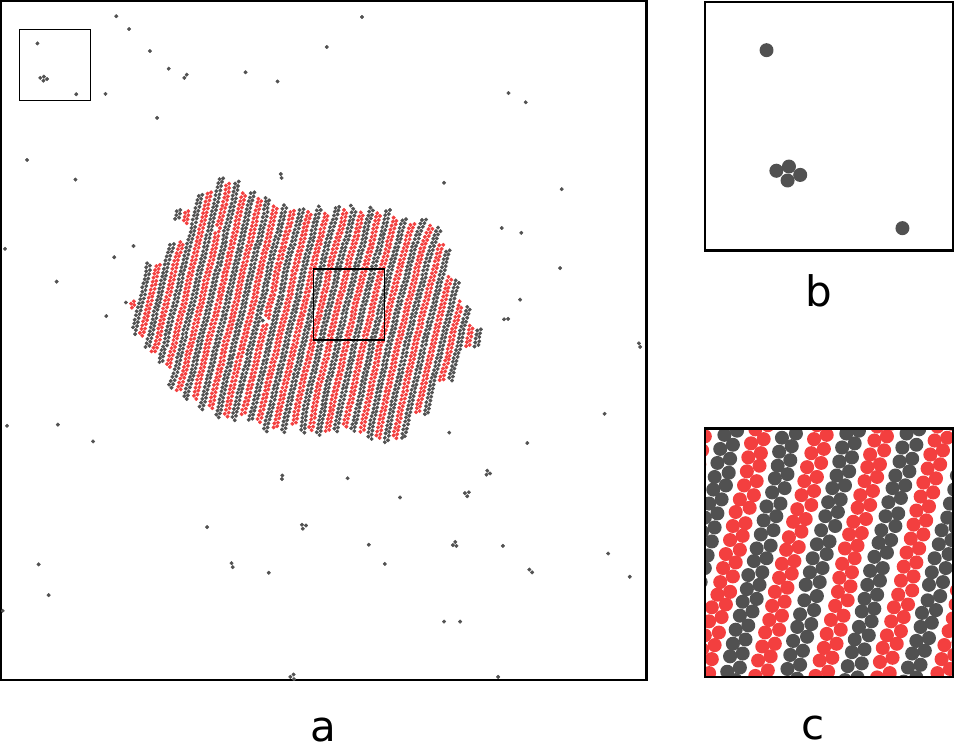}
 \caption{(a) a snapshot from MD simulations at $T^*=0.156$ for $N_1=1800$ and $N=N_1+N_2=3840$ 
  projected on the planar square with $L_x=L_y=180$. 
 The  regions inside the small frames are zoomed in (b) for the gas phase  and in  (c) for the stripe phase.
 }
 \label{fig:stripes}
 \end{figure}

\begin{figure}
 \includegraphics[scale=0.9]{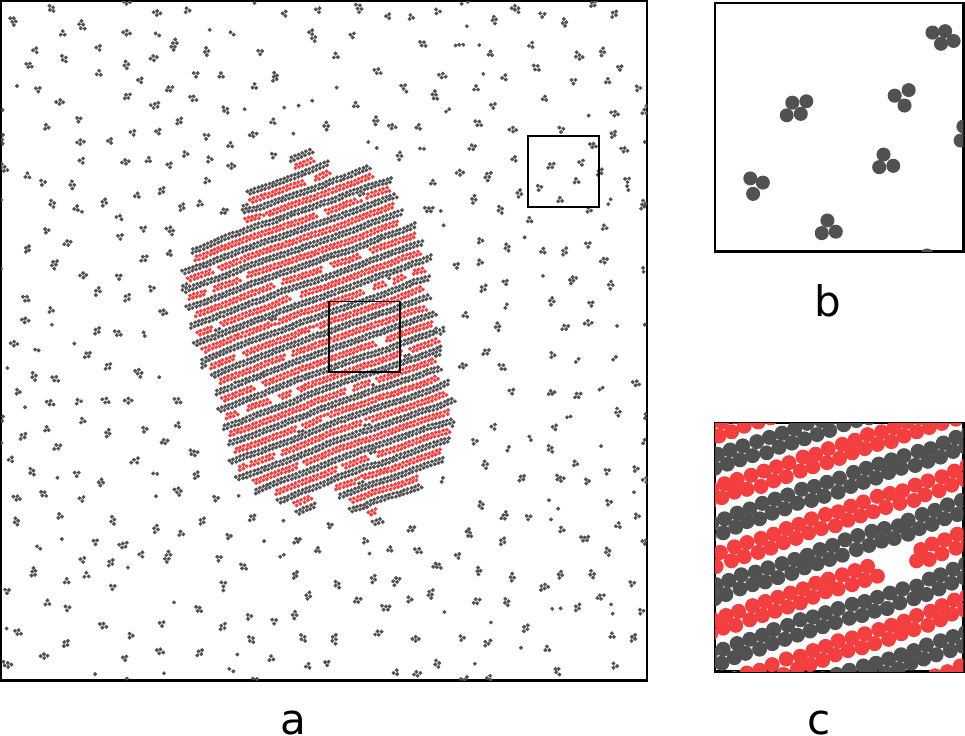}
 \caption{(a) a snapshot from MD simulations at $T^*=0.156$ for $N_1=1800$ and $N_1+N_2=5280$ projected on the planar square with $L_x=L_y=180$. The  regions inside the small frames are zoomed in (b) for the gas and in (c) for the stripe phase.}
 \label{fig:stripes1}
 \end{figure}

\begin{figure}
 \includegraphics[scale=0.9]{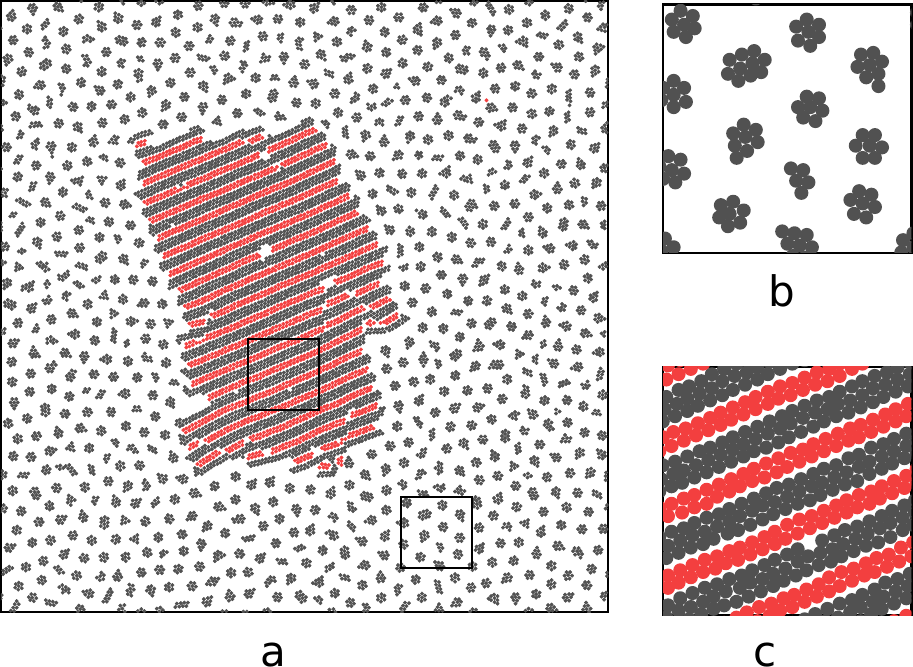}
 \caption{(a) a snapshot from MD simulations at $T^*=0.156$ for $N_1=1800$ and $N_1+N_2=10400$ projected on the planar square with $L_x=L_y=180$. The  regions inside the small frames are zoomed in (b) for the one-component clusters  and in (c) for the stripe phase. Note the larger thickness (3 particles) of the stripes of the majority component. }
 \label{fig:stripes2}
\end{figure}

\begin{figure}
\includegraphics[scale=0.9]{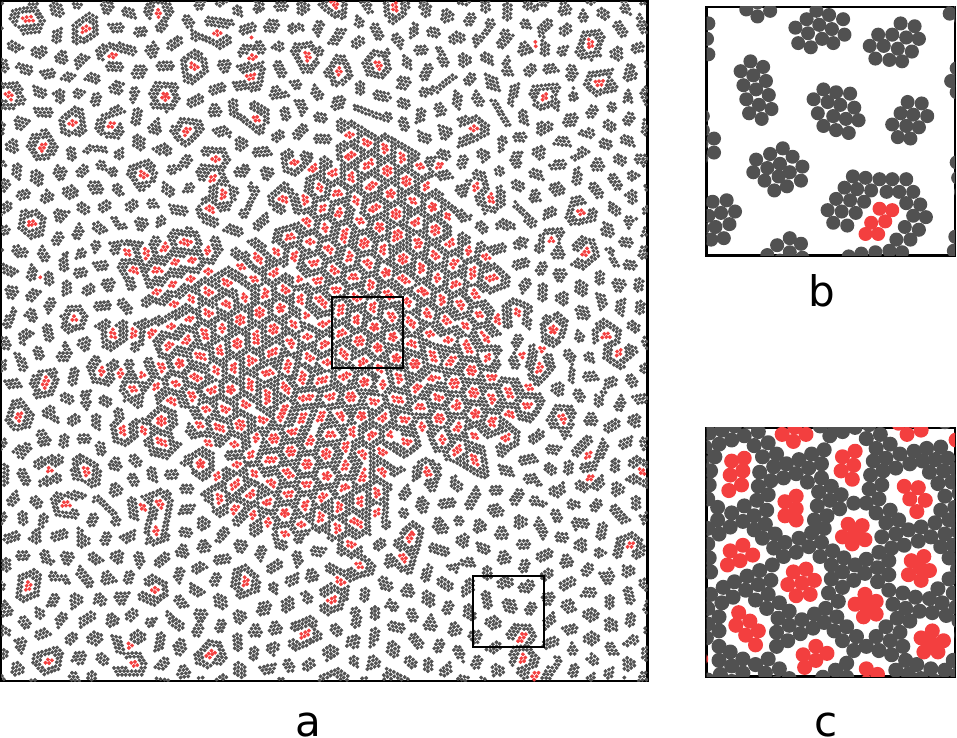}
\caption{(a) a snapshot from MD simulations at $T^*=0.156$ for $N_1=1800$ and $N_1+N_2=15200$ projected on the planar square with $L_x=L_y=180$. The  regions inside the small frames are zoomed in (b) for the one-component clusters  and in (c) for the two-component hexagonal phase. }
\label{fig:flowers1}
 \end{figure}

\begin{figure}
\includegraphics[scale=0.9]{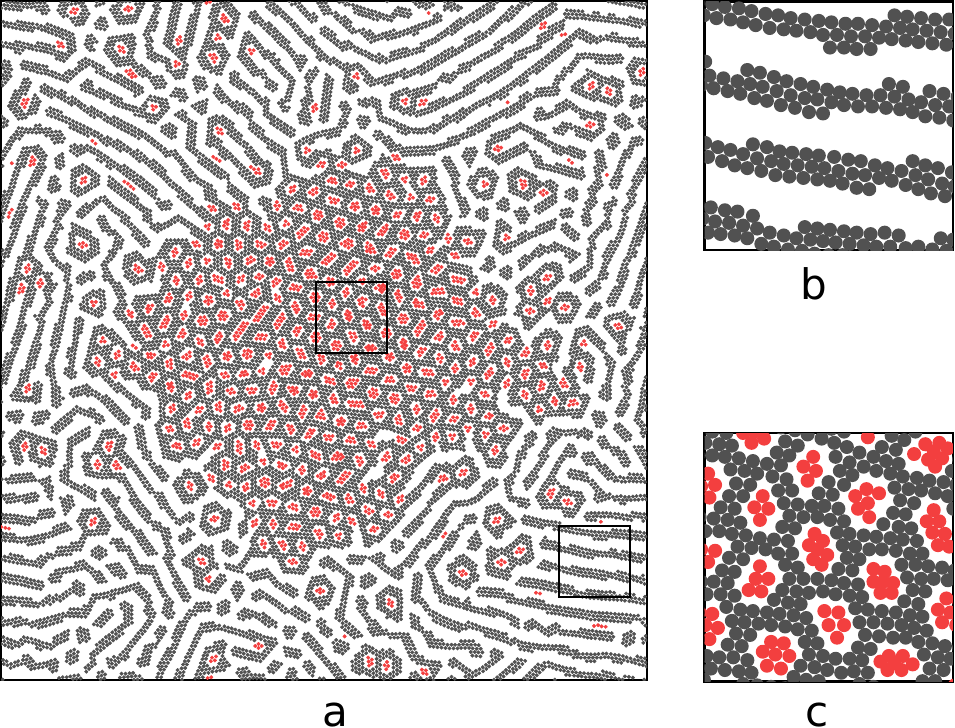}
\caption{(a) a snapshot from MD simulations at $T^*=0.156$ for $N_1=1800$ and $N_1+N_2=17600$ projected on the planar square with $L_x=L_y=180$. The  regions inside the small frames are  zoomed in (b) for the one-component stripe and in (c) for the two-component hexagonal phase.  }
\label{fig:flowers}
 \end{figure}

We simulated systems with different mole fractions  and different two-dimensional (2D) densities. For presentation of the results, we choose 5 representative cases.  Snapshots obtained for these representative systems are shown in
  Figs.\ref{fig:stripes}-\ref{fig:flowers} and the 2D densities in the coexisting phases are given in table~\ref{table}. In each selected system, $T^*=0.156$ and $N_1=1800$, whereas the total number of particles from the first to the fifth  system  
(Figs.~\ref{fig:stripes}-\ref{fig:flowers}) 
takes the value $N=N_1+N_2= 3840,5280,10400, 15200, 17600$. The average concentration 
and density both increase from Fig.\ref{fig:stripes} to Fig.\ref{fig:flowers}.

 
  
  In all the shown cases, a dense two-component solid phase coexists with a less dense fluid phase where only the majority component is present (Figs.\ref{fig:stripes}-\ref{fig:stripes2}) or the minority component forms a few dispersed clusters surrounded by shells made of the majority component (Figs.\ref{fig:flowers1} and \ref{fig:flowers}).  In Figs.\ref{fig:stripes}-\ref{fig:stripes2},
  the dense phase consists of alternating stripes of the two components, and in  Figs.\ref{fig:flowers1} and \ref{fig:flowers}, 
  clusters of the minority component occupy hexagonally distributed vacancies in the crystal of the majority component.   
  
  \begin{table}
       \hskip1.3cm    $ N$  \hskip0.65cm   $ N/L^2$  \hskip0.5cm    $\rho_s$  \hskip0.5cm  $\rho_s^1/\rho_s$  \hskip0.5cm  $\rho_f$  \hskip0.5cm $\rho_f^1/\rho_f$  \\     
   Fig4   \hskip0.55cm        3840   \hskip0.3cm   0.119    \hskip0.3cm   0.79  \hskip0.3cm     0.50  \hskip0.3cm     0.004 \hskip0.3cm    0.0000  \\  
   Fig5    \hskip0.55cm       5280 \hskip0.3cm     0.163   \hskip0.3cm    0.76    \hskip0.3cm   0.47   \hskip0.3cm    0.050  \hskip0.3cm   0.0000  \\    
   Fig6    \hskip0.4cm     10400 \hskip0.3cm     0.321  \hskip0.3cm     0.84    \hskip0.3cm   0.4   \hskip0.3cm    0.210  \hskip0.3cm   0.0005  \\  
   Fig7   \hskip0.4cm      15200  \hskip0.3cm    0.444  \hskip0.3cm     0.75   \hskip0.3cm    0.26    \hskip0.3cm   0.380  \hskip0.3cm   0.0300   \\ 
   Fig8    \hskip0.4cm     17600  \hskip0.3cm    0.543  \hskip0.3cm     0.80   \hskip0.3cm    0.26    \hskip0.3cm   0.470  \hskip0.3cm   0.0200  \\      
\caption{Systems with snapshots presented in Figs.~\ref{fig:stripes}-\ref{fig:flowers}. $N=N_1+N_2$ and $ N/L^2$ are the number of particles and the density in the quasi-2D system with area $180\times 180$, where the length unit is the particle diameter. In each case $N_1=1800$. $\rho_s$ and $\rho_f$ are the 2D densities, and  $\rho_s^1$ and $\rho_f^1$ are the 2D densities of the first component in the coexisting solid and  fluid phases, respectively.
The 2D densities in the coexisting phases were estimated by direct calculation from a central part of the crystallite, and from a large portion of the fluid phase.}
\label{table}
\end{table}

\begin{figure}
  \includegraphics[scale=0.6]{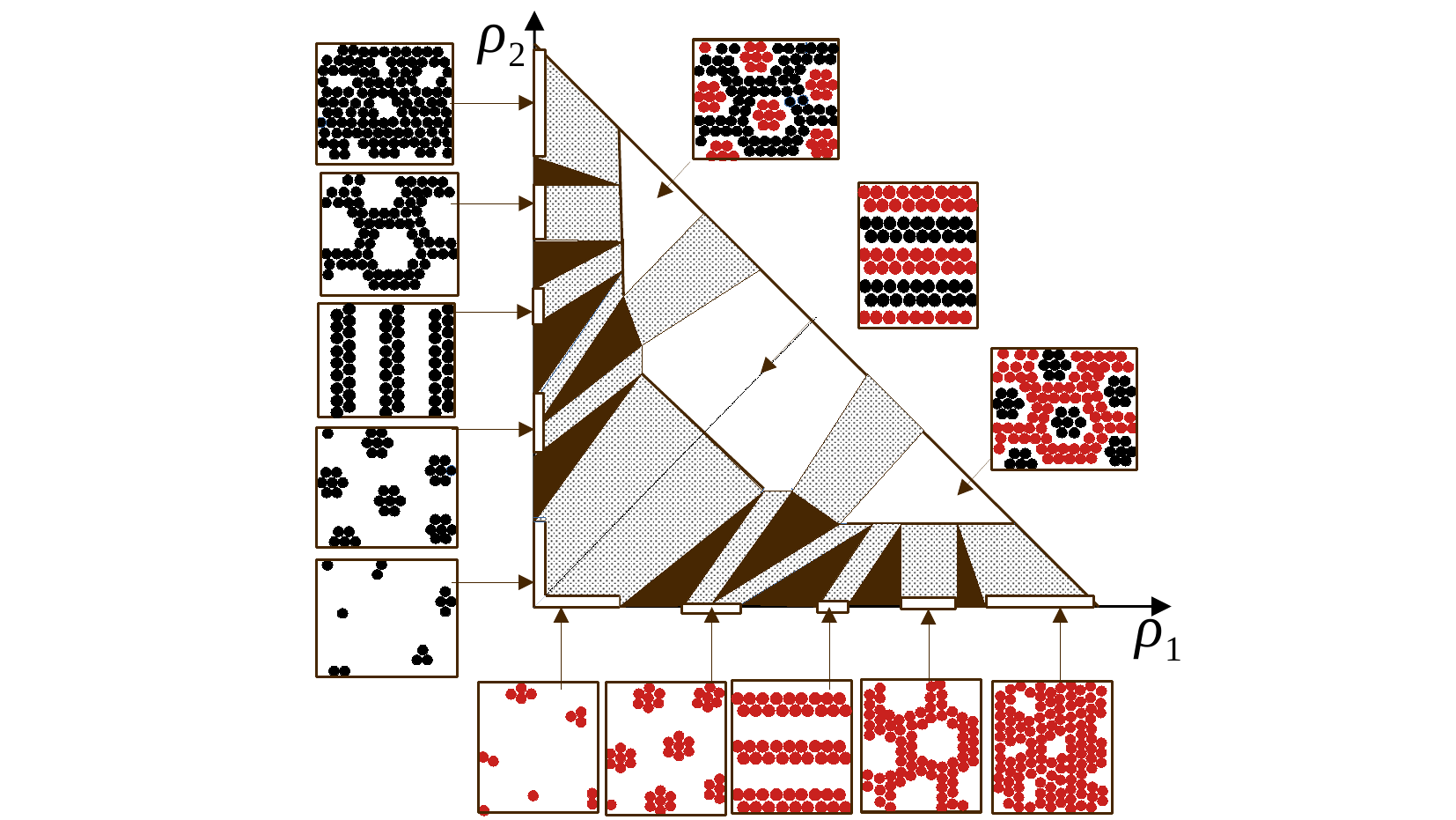 }
  \caption{Schematic phase diagram at fixed low temperature. $\rho_1$ and $\rho_2$ are the densities of the first and the second component. In the white regions, single phases are stable.
  Dark brown regions and light patterned regions  represent three-phase and two-phase equilibria, respectively. Structure of the phases 
  is illustrated in the pictures, with red and black circles representing particles of the first and the second component, respectively.
   }
  \label{fig:GS}
  \end{figure}

   We interpret all  patterns with alternating parallel stripes of the two components as the same phase denoted by S. 
  With growing difference between $N_1$ and $N_2$, first holes in the stripes of the minority component are formed (Fig.\ref{fig:stripes1}), and next stripes of the majority component become thicker (Fig.\ref{fig:stripes2}). Such evolution of the pattern leads to increased stability region of the S phase on the  
  phase diagram (Fig.\ref{fig:GS}). The mole fraction of the first component in this phase is $0.4\le N_1/N\le 0.6$ (Table~\ref{table}). In the dense phase with hexagonal symmetry, the vacancies in the majority component can be fully or partially filled with the particles of the minority component, and the mole faction of the minority component is around $1/4$.

    In order to verify if the two-component hexagonal phase is stable or metastable at low $T$, we performed additional simulations, starting from a random distribution of the particles for $\rho_1=0.19$ and  $\rho_1+\rho_2=0.76$  corresponding to the dense phase shown in Fig.~\ref{fig:flowers1}a. The resulting structure 
    was the same as the one obtained in the procedure described in sec.~\ref{sec:simulations} for the same $\rho_1$ and $\rho$, indicating that for such or similar values of the densities, the hexagonal rather than the S phase is stable at low $T$.
  
 Let  us focus on the less dense phase in  Figs.\ref{fig:stripes}-\ref{fig:flowers}. First a dilute gas of the majority component coexists with the S phase (Fig.\ref{fig:stripes}). 
When the density increases, the particles in the gas surrounding the dense phase self-assemble into clusters (Fig.\ref{fig:stripes1}). These clusters tend to form a hexagonal pattern for still larger density (Fig.\ref{fig:stripes2}).  When $N_2$ further increases with fixed $N_1=1800$,  the phase coexisting with the two-component hexagonal phase  consists of stripes of the majority component, as shown in Fig.\ref{fig:flowers} for $N=17600$. 
  For this dense system, 
  the barriers that must be overcome to reach the global minimum are very high, therefore the shown snapshot may not represent the global minimum of the free energy. 
 Similar defects were observed  in various systems self-assembling into stripes or layers Ref.~\cite{seul:95:0}.
 In Figs.~\ref{fig:flowers1} and \ref{fig:flowers}, micelle-like clusters with a core made of the minority-component surrounded by a shell consisting of the
  majority-component can be seen. They resemble structural units  of the dense phase, and at lower temperatures presumably condense on the crystallite of this phase in thermal equilibrium. The interface between the two phases is smoother in thermal equilibrium at low $T$ than in our snapshots, but we did not try to reach the final stable state, since the equilibrium shape of the monocrystal is not our goal in this study. We did not simulate still larger densities. 
  
  The simulation results and the results obtained  for the triangular-lattice model~\cite{virgiliis:24:1}, as well as the already established phase behavior of the one-component SALR systems~\cite{ciach:13:0,almarza:14:0,zhuang:16:0,archer:08:0,pini:17:0,
marlot:19:0} suggest that the low-temperature diagram should have the qualitative shape shown in Fig.~\ref{fig:GS}. The snapshots in Figs.~\ref{fig:stripes}-\ref{fig:flowers} represent evolution of the pattern along a vertical line in the diagram in Fig.~\ref{fig:GS} for  fixed $\rho_1$ smaller than in the one-phase regions.
 
    \begin{figure}
  \includegraphics[scale=0.8]{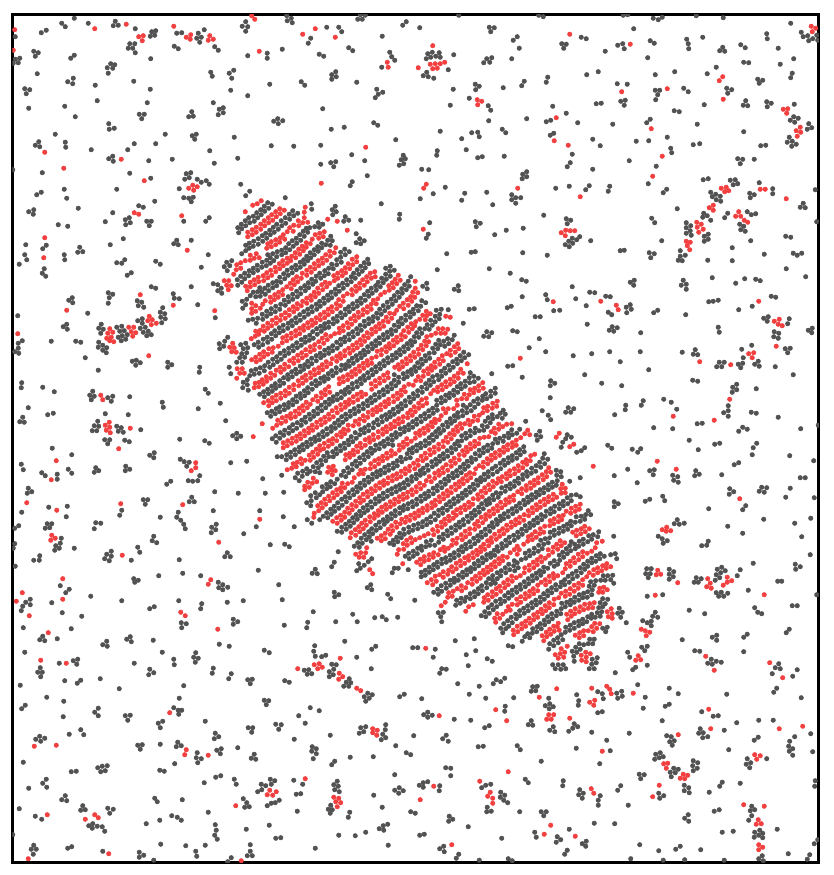}
\caption{Snapshot from MD simulations at $T^*=0.28$ for $N_1=1800, N=5280$  projected on the planar square with $L_x=L_y=180$.}
\label{fig:snap16}
 \end{figure} 

  \begin{figure}
  \includegraphics[scale=0.35]{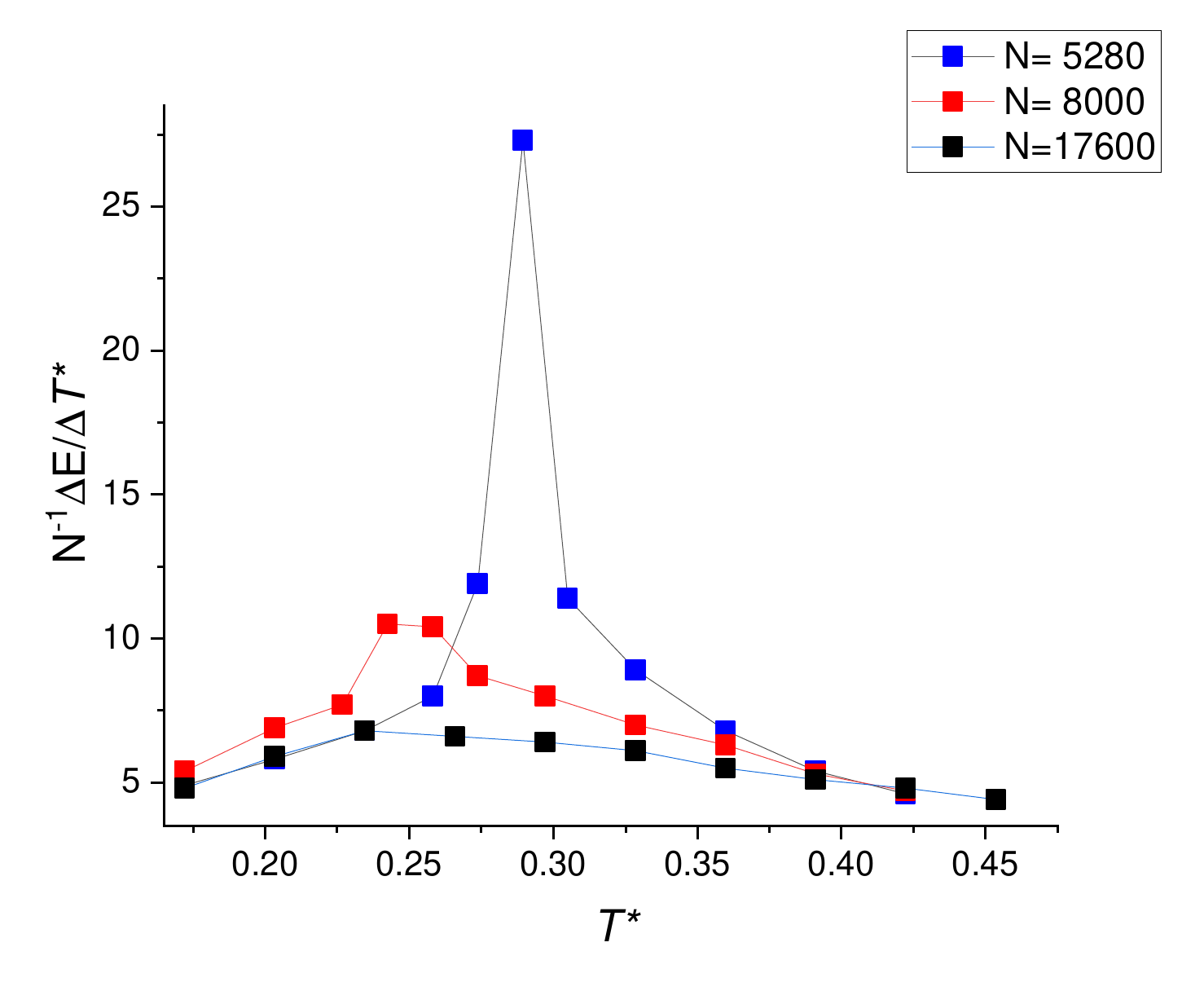}
\caption{Heat capacity per particle for $N_1=1800$ and 
$N=5280, 8000, 17600$ as a function of temperature (in $k_BT/|u_{min}|$ units). Lines are to guide the eye.}
\label{fig:cv}
 \end{figure}
   
  \begin{figure}
  \includegraphics[scale=0.35]{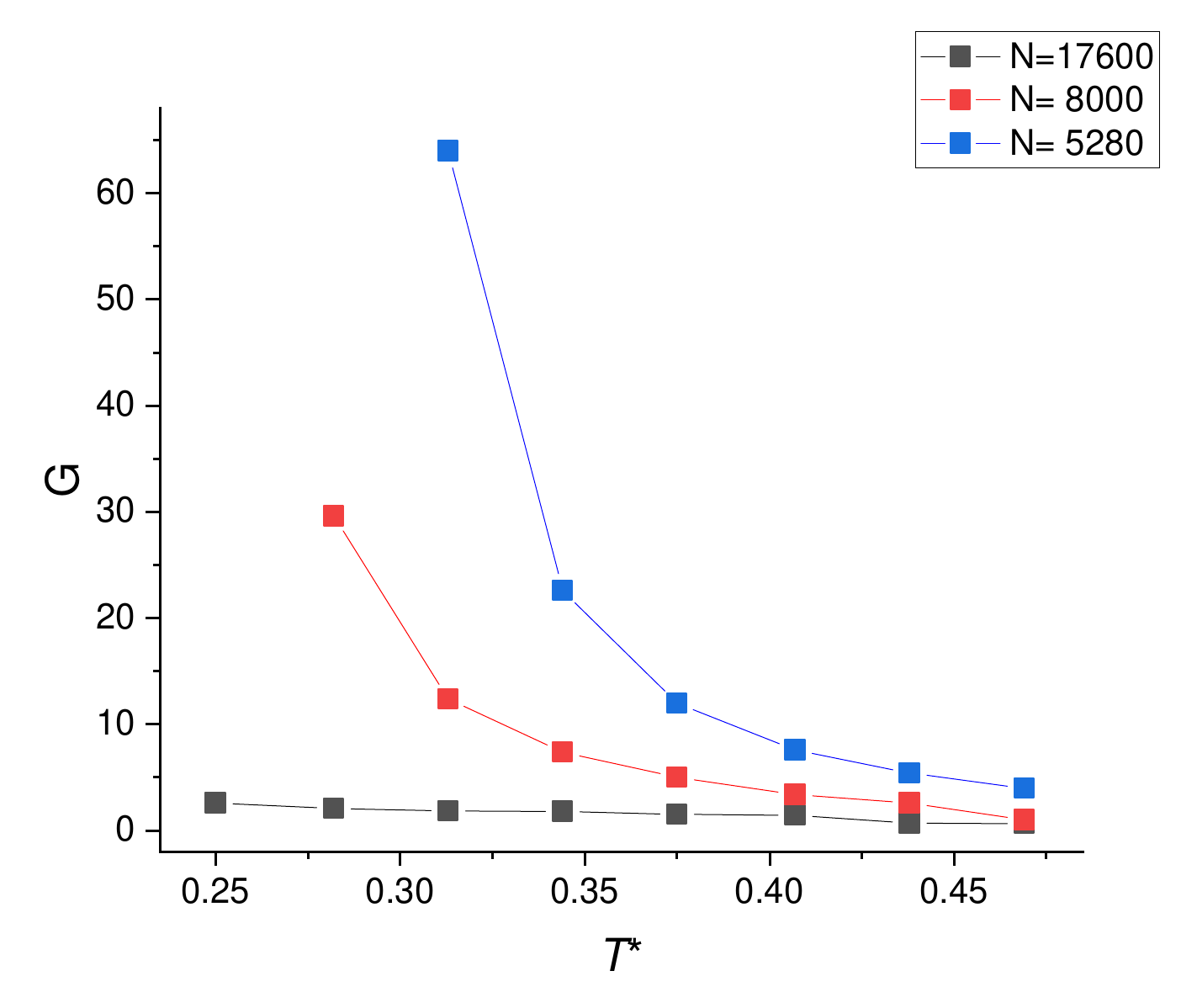}
\caption{The function $G$ equal to the integral of the correlation function and related to the compressibility according to Eq.~(\ref{G}) for $N_1=1800$ and 
$N=5280, 8000, 17600$ as a function of temperature (in $k_BT/|u_{min}|$ units). Lines are to guide the eye.}
\label{fig:corfu}
 \end{figure}
     
    The one-component ordered phases loose stability at relatively low $T$ (see Fig.~\ref{f3}). Above their melting temperature, the dense two-component phases coexist  with the disordered phase, and at this coexistence 
  become less ordered and less dense (see Fig.~\ref{fig:snap16} for $T^*=0.28$).  The structure of the disordered phase, however,  becomes more complex and self-assembly into large two-component super-clusters takes place.
    Further heating leads to melting of the periodic patterns made by the two components. 
    
    To study the transitions to the disordered phase, we considered specific heat and compressibility as functions of $T$.  
     The specific heat $C_V=N^{-1}d \langle E\rangle/dT$ is shown in Fig.~\ref{fig:cv} with the derivative approximated by $\Delta E/\Delta T$, with $\Delta T=0.0313$ or $\Delta T=0.0156$ far from or close to the maximum, respectively. In Fig.~\ref{fig:corfu} we show the integral of the correlation  function,
\begin{equation}
\label{G}
G=2\pi\int_0^{\infty} (g(r)-1)r dr=k_BT \chi_T-1/\rho,
\end{equation} 
where $\chi_T$ is the compressibility. Since we consider a quasi-two dimensional system,  we used the two-dimensional integral in (\ref{G}).
    
 The maxima of $C_V$ and $G$  in  Figs.~\ref{fig:cv} and \ref{fig:corfu} suggest a phase transition for $\rho_1/\rho\approx 0.34$ and $\rho_1/\rho\approx 0.25$,
   but  for $\rho_1/\rho\approx 0.1$ no pronounced maxima can be seen in the shown temperature interval.   When
    $\rho_1/\rho\approx 0.34$, the transition occurs at higher temperature  than for $\rho_1/\rho\approx 0.25$.
   These results agree with theoretical expectations of increasing temperature interval of the stability of the ordered phase with decreasing $|\rho_1-\rho_2|$  (see Fig.~\ref{f3}).
 
 \subsection{Structure of the disorderd phase}
  \label{sec:results2}
  
\begin{figure}
 \includegraphics[scale=0.8]{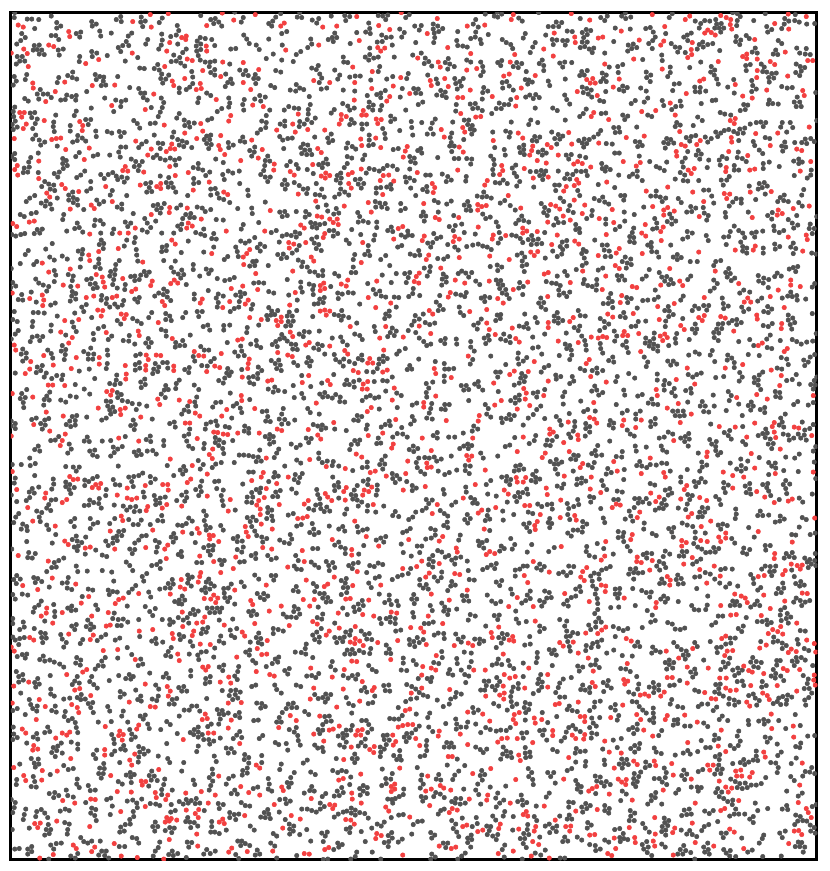}
  \caption{Snapshot from MD simulations at $N_1=1800, N=8000$, $T^* = 0.438 $ projected on the planar square with $L_x=L_y=180$.}
  \label{fig:snapdisl}
\end{figure}

\begin{figure}
  \includegraphics[scale=0.8]{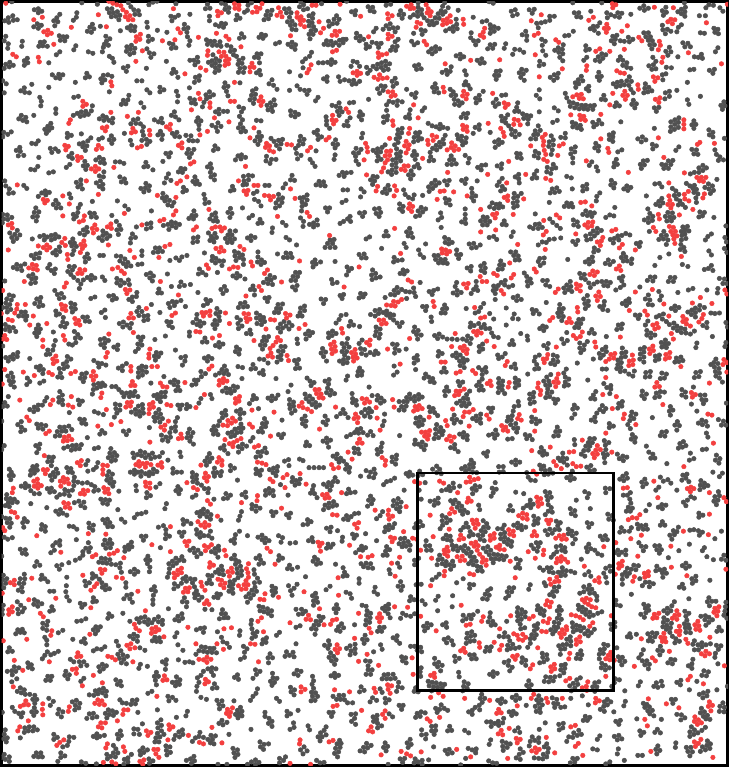}  
  \includegraphics[scale=0.8]{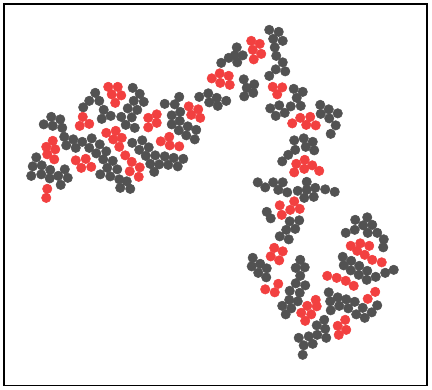}
  \vfill a \hskip 8 cm b
  \caption{(a) a snapshot from MD simulations for $N_1=1800, N=8000$ and $T^* = 0.313 $ projected on the planar square with $L_x=L_y=180$. 
 Note the super-cluster consisting of $280$ particles inside the small frame. (b) The above super-cluster zoomed in.}
  \label{fig:snapdis}
\end{figure}

\begin{figure}
\includegraphics[scale=0.8]{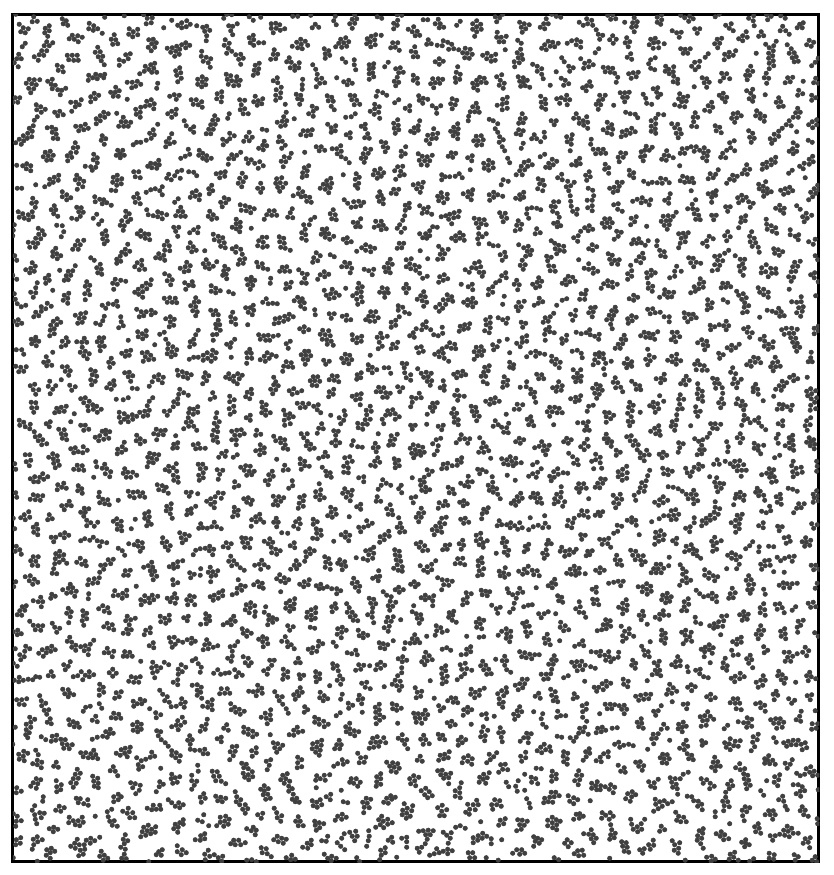} 
  \caption{Snapshot from MD simulations at $T^* = 0.313 $  for $N_1=0$, $N=8000$ (one-component SALR system) projected on the planar square with $L_x=L_y=180$. Total number of particles and $T^*$ are the same as in Fig.~\ref{fig:snapdis}.}
  \label{fig:snapone}
\end{figure}

  Characteristic snapshots in the disordered phase are shown  for  $N_1=1800$ and $N=8000$  in Figs.~\ref{fig:snapdisl} and~\ref{fig:snapdis} for $T^*=0.438$  and $T^*=0.313$, respectively. In addition to single particles of each component
   and small clusters of the majority component, aggregates of alternating one-component clusters of various sizes are present. An example of such an aggregate that we call a super-cluster is shown in Fig.~\ref{fig:snapdis}. For comparison, in Fig.~\ref{fig:snapone} we show a snapshot of a one-component system ($N_1=0$) with the same $N$ and $T^*$ as in Fig.~\ref{fig:snapdis}.
  One can see  by visual inspection  that in the one-component SALR system the size polydispersity of the aggregates is much smaller than
   in our mixture, and no clusters as large as the super-cluster shown in Fig.~\ref{fig:snapdis} are present.

\begin{figure}
   \includegraphics[scale=0.315]{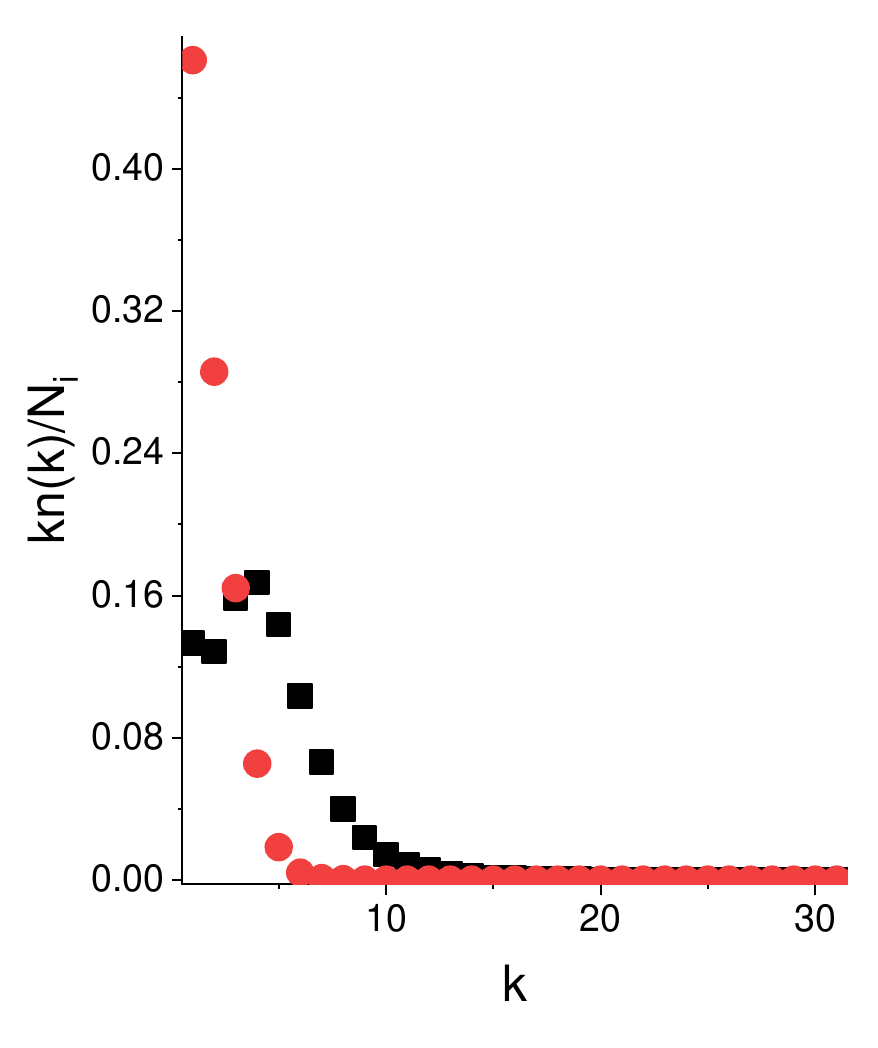} 
    \includegraphics[scale=0.315]{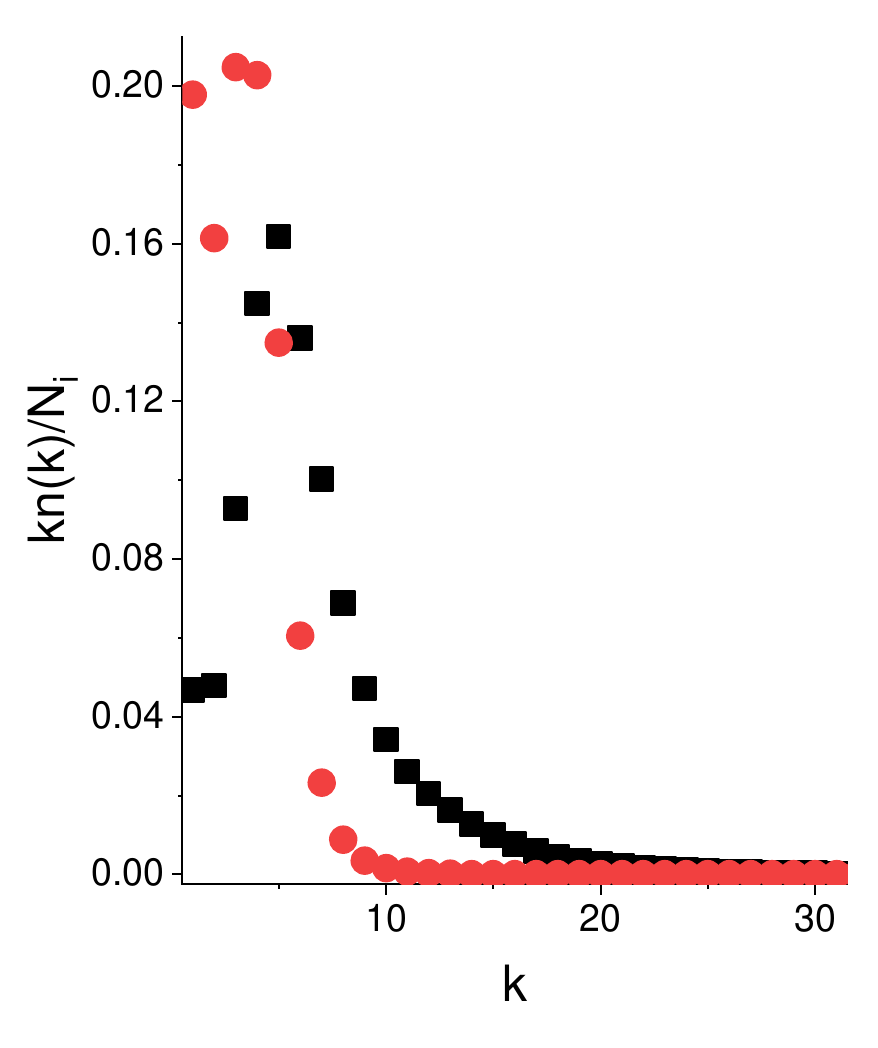}
  \includegraphics[scale=0.315]{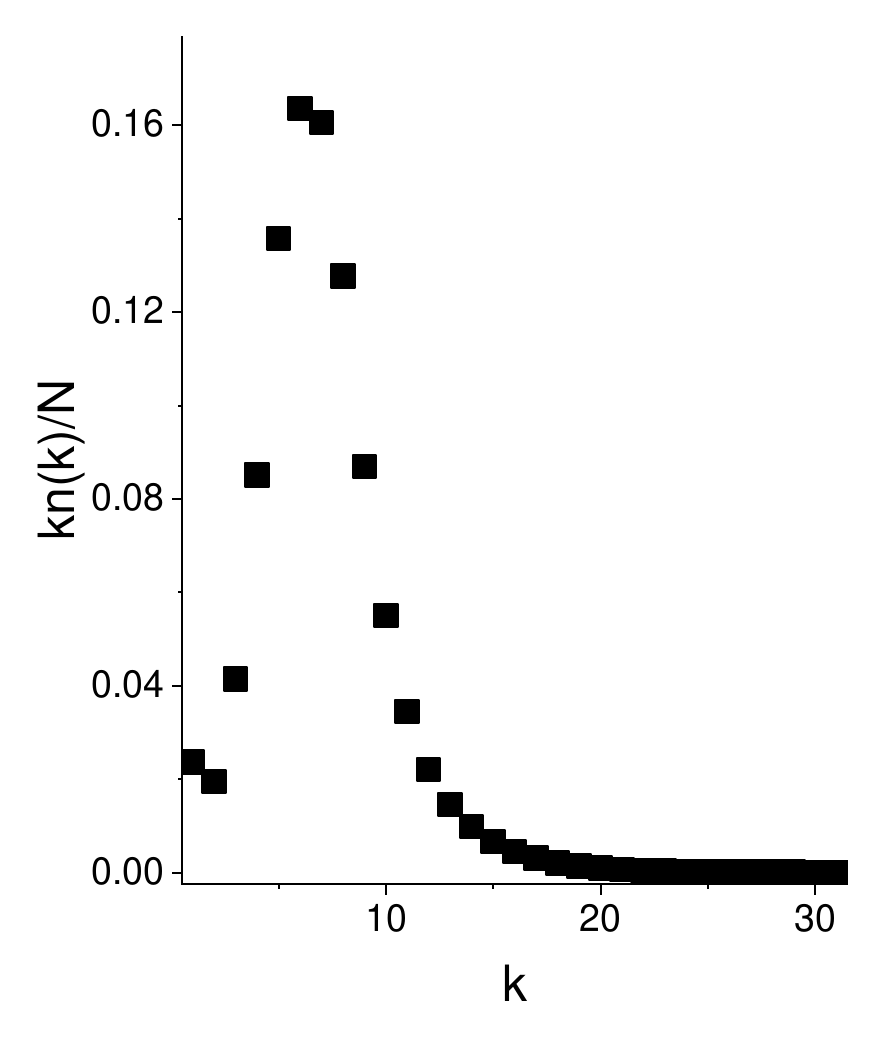}
 \vfill a\hskip 5cm b\hskip 5cm c
  \caption{The number of particles, $kn(k)$, belonging to a one-component cluster consisting of $k$ particles, divided by
 the number $N_i$  of the particles of this component in the system. In each case, $ N=N_1+N_2=8000$  and (a) $N_1=1800$,  $T=0.438$, (b)  $N_1=1800$ and $T=0.313$, and (c) $N_1=0$,  and $T=0.313$. Red and black symbols correspond to the minority ($i=1$)  and the majority ($i=2$) component, respectively. }
  \label{fig:kn(k)}
\end{figure}

 To quantify the size distribution of the clusters, we first consider each component separately and compute the number of clusters consisting of k particles, $n(k)$, as well as  $kn(k)$, i.e. the number of  particles of the considered component belonging to a cluster consisting of $k$ particles of the same component. 
 $kn(k)$  is presented for  $N_1=1800, N=8000$ and two temperatures, $T^* = 0.438, 0.313 $, for each component in Fig.~\ref{fig:kn(k)}a and Fig.~\ref{fig:kn(k)}b. For comparison, we show  in Fig.~\ref{fig:kn(k)}c $kn(k)$ for $T^* = 0.313$,  $N=8000$ and $N_1=0$.  The snapshots corresponding to the histograms in Fig.~\ref{fig:kn(k)}a,b,c are shown in Figs.~\ref{fig:snapdisl}, \ref{fig:snapdis} and ~\ref{fig:snapone}, respectively.
 
  At the higher $T^*$, single particles of the minority component dominate,  $kn(k)$ decreases from the maximum at $k=1$, and clusters larger than $k=6$ are absent. On the other hand, particles of the majority component are most probably members of 5-particle clusters, and clusters consisting of up to 15 particles can be seen. At the lower  $T^*$, self-assembly of the minority component into 3- or 4-particle clusters is more probable than formation of dimers or monomers, $kn(k)$ has a minimum at $k=2$ and the largest (rare) clusters consist of 8 particles.  While the histograms of the minority component change with decreasing $T$ qualitatively,  the histograms of the majority component remain similar, except that 
 much fewer particles of the majority component remain in the form of monomers at the lower temperature. By comparing Fig.~\ref{fig:kn(k)}b with Fig.~\ref{fig:kn(k)}c, we can see similarity between the distribution of the particles of the majority component with the distribution of the particles in the one-component system. 
 
 Similarity between the histograms for the second component in Fig.~\ref{fig:kn(k)}b and c  disagrees with the visual impression of significantly different distribution of the particles in Figs.~\ref{fig:snapdis} and \ref{fig:snapone}.
This is because the distribution of the one-component clusters does not give enough information about their connectivity, clearly seen in the snapshots. Unfortunately,
there is no unique definition of the two-component super-cluster, because there is no unique way of defining the connectivity between the compact one-component clusters, when the minimum of the energy of a pair of different particles is  very flat (Fig.\ref{f1}). We  choose for the analysis the presence of at least four bonds between  particles belonging to different one-component clusters as the criterium that these clusters belong to a super-cluster. For the details of the super-cluster definition see  sec.~\ref{sec:simulations}. We use the same notation, $n(k)$, for the number of super-clusters consisting of $k$ particles.  

\begin{figure}
 \includegraphics[scale=0.32]{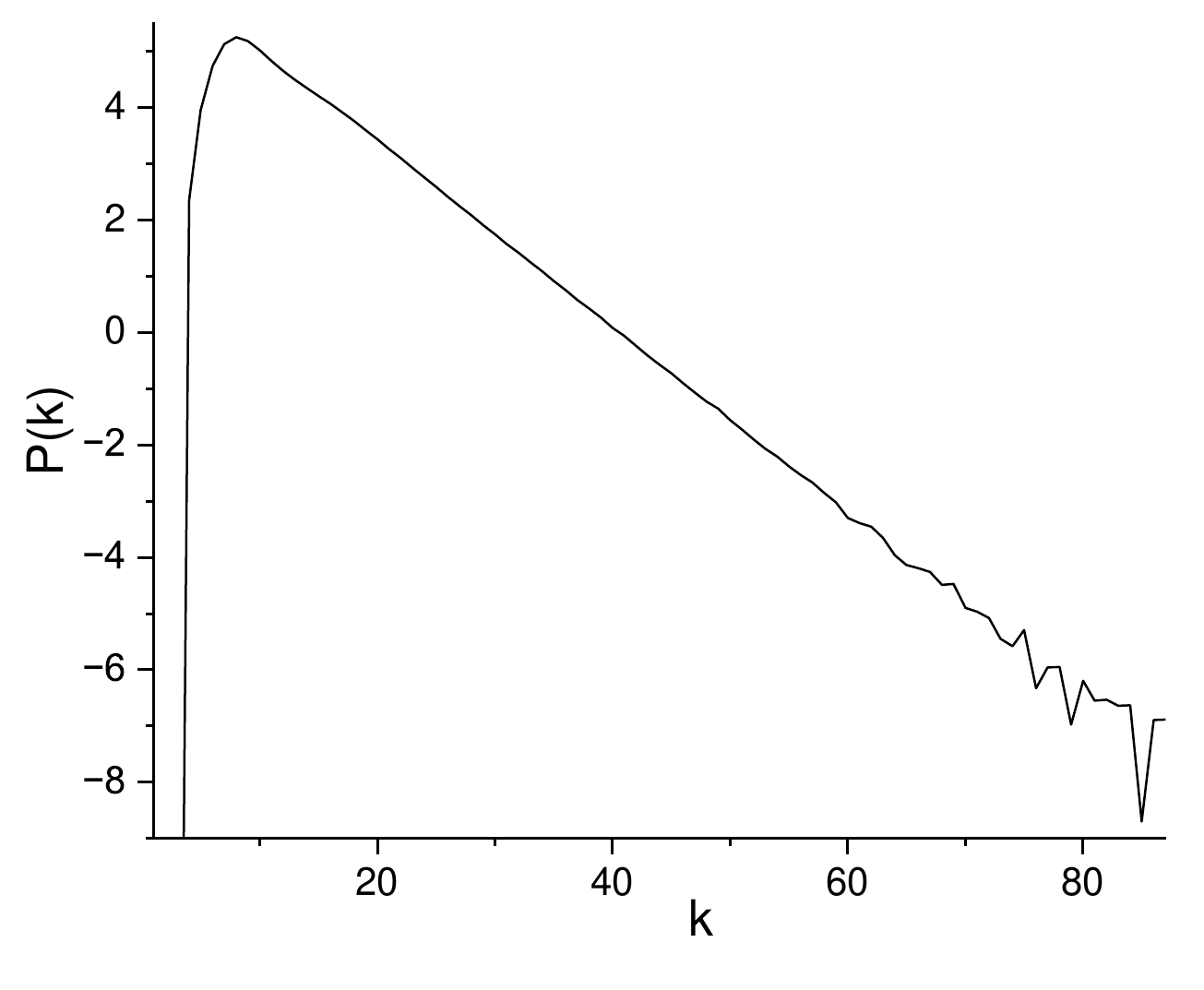}
   \includegraphics[scale=0.32]{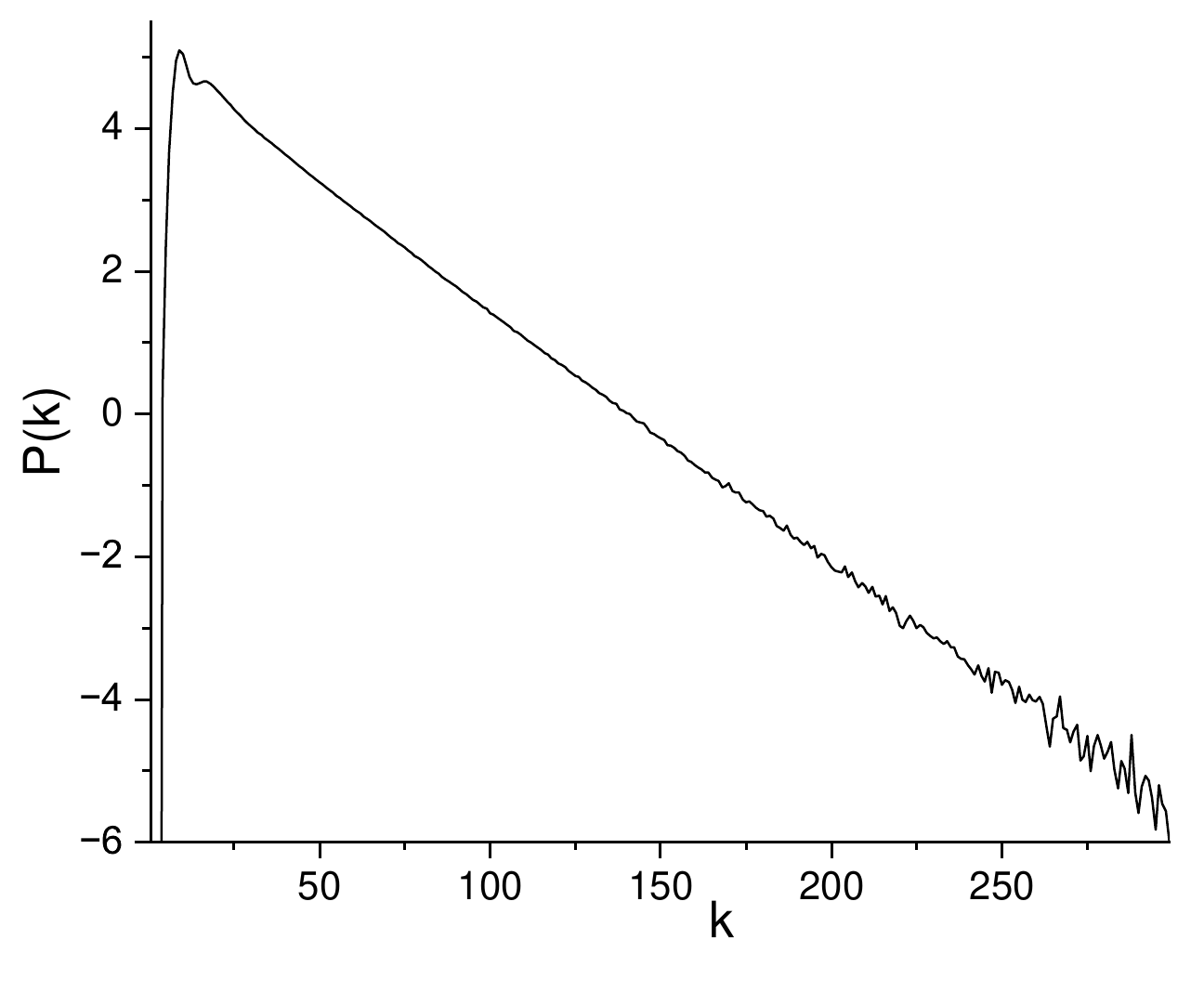} 
   \vfill a \hskip 8cm b
  \caption{$P(k)$ (Eq.~(\ref{P(k)})) for $N_1=1800$ and $N=8000$. (a) $T^*=0.438$ and (b)  $T^*=0.313$.}
  \label{fig:p(n)}
\end{figure}

In order to  analyse the super-clusters distribution, in
 Fig.~\ref{fig:p(n)} we plot the function

\begin{equation}
  P(k)= \ln(kn(k))
  \label{P(k)}
  \end{equation}
for the same thermodynamic states as in Fig.~\ref{fig:kn(k)} a, b. Linear dependence of $P(k)$ on $k$ is clearly seen for suffciently large $k$ in Fig.~\ref{fig:p(n)}, suggesting that $n(k)$ has the following functional form 
 
 \begin{equation}
 \label{p1(n)}
 n(k)\propto \frac{ \exp(-\alpha k)}{k}, 
 \end{equation}
and that the probability of finding a particle in the super-cluster consisting of $k$ particles (proportional to $kn(k)$) has
the exponential form, 
  \begin{equation}
 \label{p(n)}
 p(k)= A\exp(-\alpha k),
 \end{equation}
  where $A$  is the normalization constant.  
  $A\approx \alpha$ for integer $k$ when $\alpha\ll 1$. The exponential form means that
   $p(k)=p(k_1)p(k-k_1)/A$ for any  $k_1\le k/2$, i.e. a formation of a large super-cluster occurs with a probability that is approximately a product of probabilities of findig  super-clusters consisting of $k_1$ and $k-k_1$ particles, times $\alpha^{-1}\gg 1$. Since $1/\alpha$ roughly measures the size of the largest super-clusters, for $k\sim1/\alpha$ we have $p(k)\sim\sum_{k_1=1}^{k/2} p(k_1)p(k-k_1)$.
  Thus, the large super-clusters appear and disappear by forming and breaking bonds between smaller super-clusters almost randomly.
   
   We should note that $P(k)$ depends on the definition of the connectivity between the cluster quantitatively, but not qualitatively, i.e. we get the linear dependence of $P(k)$ on $k$ with different slopes for different connectivity criteria. Importantly, for given connectivity criterium $\alpha$ takes the same value for different sufficiently large values of $N$. We obtained the same results for $\alpha$ for $N=8000$ and $N=32 000$.

\begin{figure}
  \includegraphics[scale=0.5]{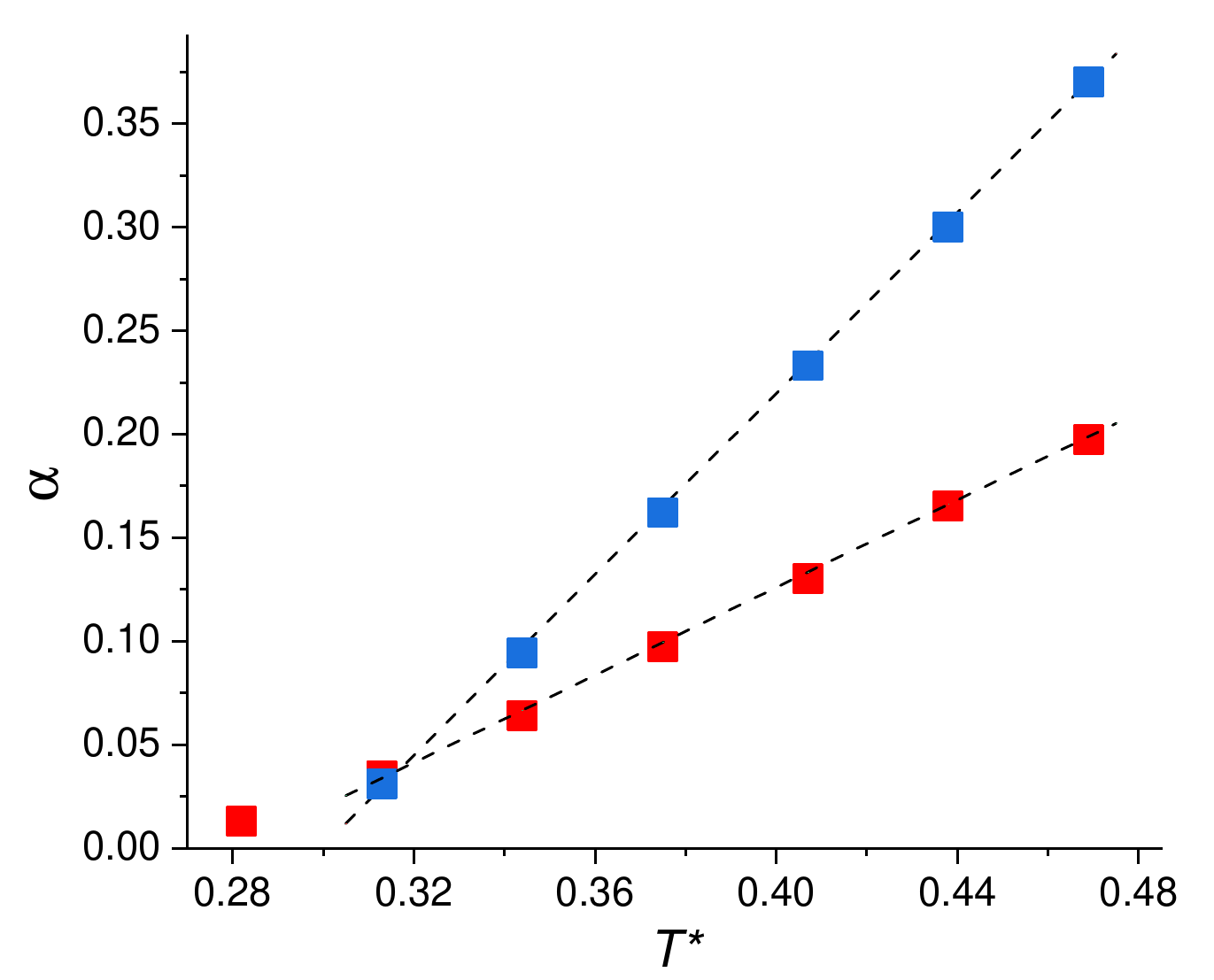}
\caption{The parameter $\alpha$ in Eq.~(\ref{p(n)}) for $N=8000$ (blue symbols) and 
$N=5280$ (red symbols) as a function of temperature $T^*=k_BT/|u_{min}|$. Dashed lines are the stright lines fitted to our reults.}
\label{fig:alpha}
 \end{figure}

\begin{figure}
  \includegraphics[scale=0.45]{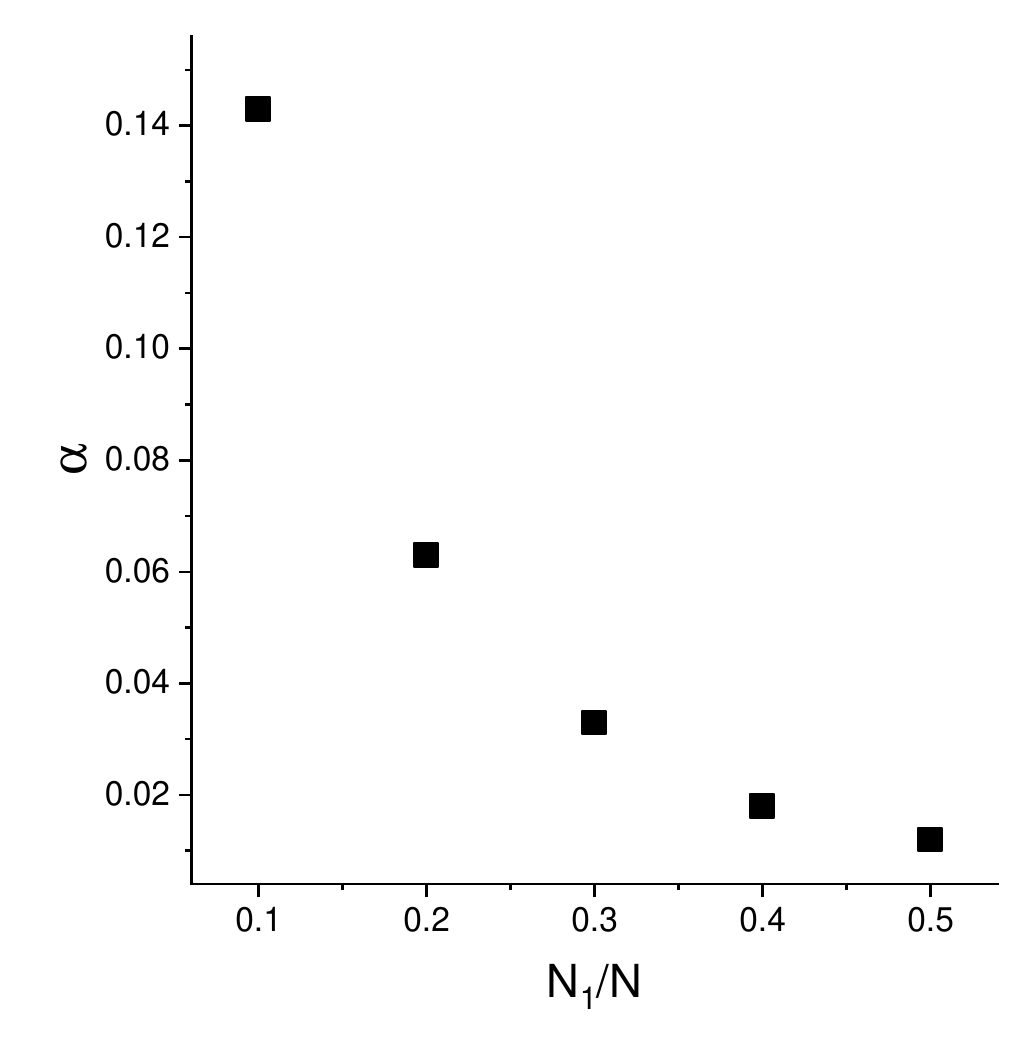}  \includegraphics[scale=0.45]{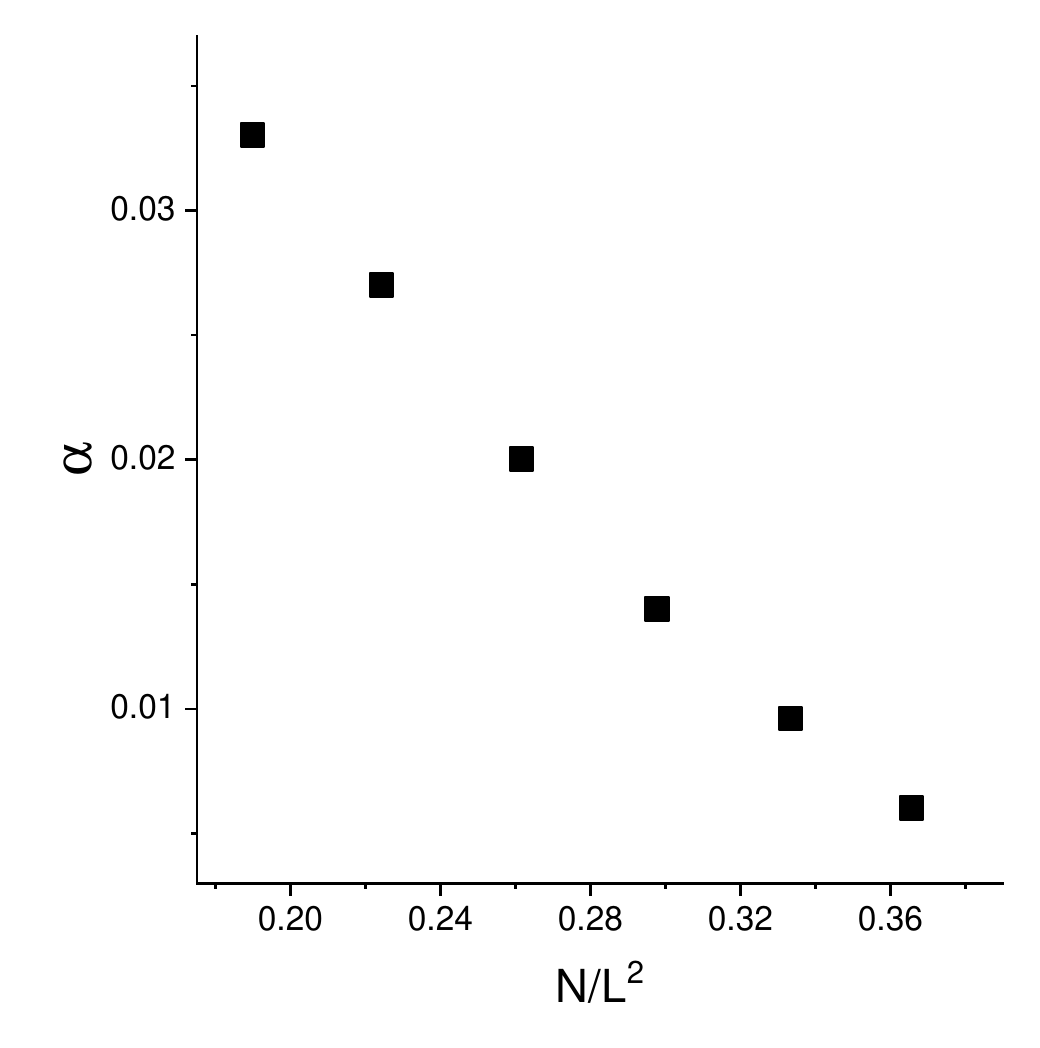}
\caption{The parameter $\alpha$ in Eq.~(\ref{p(n)})  as a function of the mole fraction $N_1/N$ for $N=8010$ ($N/L^2=0.19$) and $T^*=0.313$ (left) and   as a function of
$N/L^2$ for the mole fraction $N_1/N=0.3$ and $T^*=0.313$ (right).}
\label{fig:alpha1}
 \end{figure}

 The size of the largest super-clusters is of order of $1/\alpha$ that depends on the thermodynamic state.  
 We considered a number of systems with different $T^*$, $N$ and $N_1$, and determined $\alpha$ from the slopes of $P(k)$.  
  $\alpha$  is shown  in Fig.~\ref{fig:alpha} for  $N_1=1800$ and $N=5280, 8000$ as a function of temperature, and in Fig.~\ref{fig:alpha1} for  $T^*=0.313$  as a function of the molar fraction $N_1/N$ for fixed number of particles  $N=8010$, and  as a function of
$N/L^2$ for the mole fraction $N_1/N=0.3$. The results of simulations (symbols) perfectly fit the stright lines in Fig.~\ref{fig:alpha}, clearly indicationg the 
linear dependence of $\alpha$ on $T^*$.
 
\section{Conclusions}
\label{sec:conclusions}

We investigated patterns with long- and short-range order formed spontaneously at low and high temperature, respectively, in a monolayer of a binary mixture of particles or macromolecules. In the studied generic model, alternating stripes of the two components are energetically favoured. We did not determine the full $(\rho_1,\rho_2,T^*)$ phase diagram, but from our present and from the previously obtained results, we can draw general conclusions concerning evolution of the pattern formation. 

The phase coexistences expected at very low temperature are shown schematically in Fig.~\ref{fig:GS}.  
The ordered dense phases of microsegregated components consist of alternating stripes of the two components or of clusters of the minority component filling the hexagonally distributed voids in the crystal of the majority component. These two-component phases coexist with the phases formed by the majority component. For increasing density of the majority component, the  phases coexisting with the two-component crystals have the structure predicted previously for the one-component SALR systems (Figs.~\ref{fig:stripes}-\ref{fig:flowers}). We expect such phase behavior for various mixtures with energetically favored alternating thin stripes of the two components~\cite{virgiliis:24:1}. When the interactions favor much wider stripes, however, at low temperatures the microsegregation into stripes takes place, and  only at intermediate temperatures the hexagonal pattern occurs~\cite{patsahan:24:0}. Thus, the phase diagram depends significantly on the width of the self-assembled aggregates that in turn depens on the shape of the interactions.

The one-component ordered phases melt at relatively low temperature upon heating the system (Fig.~\ref{f3}), and the two-component dense phases coexist
with the disordered phase. At the coexistence with the disordered phase, the phases with periodically distributed components become less dense and less ordered, and the structure of the disordered phase becomes quite complex  (Fig.~\ref{fig:snap16}). Further heating leads to transition  between the hexagonal and disordered phases, and finally between the stripe and the disordered phases (Fig.~\ref{fig:cv} and \ref{fig:corfu}).  

The snaphots of the disordered phase (Figs.~\ref{fig:snap16},   \ref{fig:snapdisl} and \ref{fig:snapdis}) show a complex pattern that looks like a rather chaotic mixture of individual particles, one-component clusters and two-component super-clusters with various sizes. The super-cluster size distribution, however, is surprisingly simple, $kn(k)\propto\exp(-\alpha k)$, where $n(k)$ is the average number of  super-clusters consisting of $k$ particles. Moreover, the parameter $\alpha$ is proportional to temperature (Fig.~\ref{fig:alpha}), i.e. the size of the largest super-clusters, $\sim 1/\alpha$, is inversely proportional to temperature. 
Super-clusters of different sizes behave like different components in a multicomponent mixture, and the equilibrium between assembly and disassembly of the super-clusters resembles equilibrium in chemical reactions in such a mixture.
 
From the fundamental point of view, it is important  that only two types of ordered patterns
with laterally microsegregated components  can appear in monolayers of the considered  binary mixtures.
 While the symmetries of the  2D ordered phases in the one-component SALR and in our mixture are the same, the structure of the disordered phase in the two cases is significantly different. It is intriguing that the size distribution of the super-clusters in the considered
 mixture  has a very simple exponential form. 
 
 From the  point of view of potential applications, it is important that it should be  easier to obtain spontaneously ordered patterns in this type of mixtures 
than in the absence of the second component, due to the much larger temperature range of stability of the phases with the long-range order. 

\section*{Conflicts of interest}
There are no conflicts to declare.
\section*{Data availability}
The data that support the findings of this study are available within the article  and from the corresponding author upon reasonable request.

\section*{Acknowledgments}
We gratefully acknowledge the financial support from the European Union Horizon 2020 research 
and innovation programme under the Marie
Sk\l{}odowska-Curie grant agreement No 734276 (CONIN).


  \end{document}